\documentclass[preprint,aps,showpacs,tightenlines]{revtex4-1}
\usepackage{amsmath}
\usepackage{epsfig}
\begin{document}

\newcommand{\beq}{\begin{eqnarray}}
\newcommand{\eeq}{\end{eqnarray}}

\newcommand{\non}{\nonumber\\ }
\newcommand{\ov}{\overline}
\newcommand{\rmt}{ {\rm T}}
\newcommand{\psl}{ P \hspace{-2.5truemm}/ }
\newcommand{\nsl}{ n \hspace{-2.2truemm}/ }
\newcommand{\vsl}{ v \hspace{-2.2truemm}/ }



\def \ctp{Commun.Theor.Phys.  }
\def \epjc{ Eur. Phys. J. C }
\def \jpg{  J. Phys. G }
\def \npb{  Nucl. Phys. B }
\def \plb{  Phys. Lett. B }
\def \pr{  Phys. Rep. }
\def \rmp{ Rev. Mod. Phys. }
\def \prd{  Phys. Rev. D }
\def \prl{  Phys. Rev. Lett.  }
\def \zpc{  Z. Phys. C  }
\def \jhep{ J. High Energy Phys.  }
\def \ijmpa { Int. J. Mod. Phys. A }


\title{The semileptonic decays $B/B_s \to (\pi, K)(l^+l^-,l\nu,\nu\bar{\nu} )$
in the perturbative QCD approach beyond the leading-order}
\author{Wen-Fei Wang, and  Zhen-Jun Xiao\footnote{Email Address: xiaozhenjun@njnu.edu.cn}}
\affiliation{ Department of Physics and Institute of Theoretical Physics,\\
Nanjing Normal University, Nanjing, Jiangsu 210046, People's Republic of China}
\date{\today}
\begin{abstract}
In this paper we first calculate the form factors of $B \to (\pi,K)$ and $B_s \to K $ transitions
by employing the perturbative QCD (pQCD) factorization approach with the inclusion of the next-to-leading-order(NLO)
corrections, and then we calculate the branching ratios of the corresponding semileptonic decays
$B/B_s \to (\pi, K)(l^+l^-,l\nu,\nu\bar{\nu} )$ (here $l$ denotes $e, \mu$ and $\tau$).
Based on the numerical calculations and phenomenological analysis, we found the following results:
(a) For $B \to (\pi, K)$ and $B_s \to K$ transition form factors $F_{0,+,\rmt}(q^2)$, the
NLO pQCD predictions for the values of $F_{0,+,\rmt}(0)$ and their $q^2$-dependence agree well with those
obtained from other methods;
(b) For $\bar{B}^0 \to \pi^+l^-\bar{\nu}_l, \bar{K}^0 l^+ l^- $ and $B^- \to \pi^0 l^-\bar{\nu}_l, K^- l^+ l^-$
decay modes, the NLO pQCD predictions for their branching ratios agree very well with the measured values;
(c) By comparing the pQCD predictions for $Br(\bar{B}^0 \to \pi^+l^-\bar{\nu}_l)$ with the measured decay
rate we extract out the magnitude of $V_{ub}$: $|V_{ub}|= \left ( 3.80^{+0.56}_{-0.50}(theor.) 
\right) \times 10^{-3}$;
(d) We also defined several ratios of the branching ratios, $R_\nu, R_C$ and $R_{N1,N2,N3}$, and presented
the corresponding pQCD predictions, which will be tested by LHCb and the forthcoming Super-B experiments.
\end{abstract}


\maketitle

\section{Introduction}\label{sec:1}

The semileptonic decays $B \to (\pi, K)(l^+l^-,l\bar{\nu}, \nu \bar{\nu})$ and $B_s \to K (l^+l^-,l\bar{\nu}, \nu \bar{\nu})$
with $l=(e,\mu,tau)$ are very interesting $B/B_s$ decays modes and playing an important role in
testing the standard model (SM) and in searching
for the new physics (NP) beyond the SM, such as the determination of $|V_{ub}|$
and the extractions of the transition form factors $F_{0,+,\rmt}(q^2)$ of $B/B_s$ meson
to pion and/or kaon. For  the charged current $B/B_s \to P l\bar{\nu}$ decays, the "Tree" diagrams provide the leading order
contribution. For the neutral current $B/B_s \to P l^+ l^-$ and $P\nu \bar{\nu}$ decays, however,
the leading order SM contributions come from the photon penguin, the Z penguin and
the $W^+W^-$ box diagrams, as shown in Fig.~1, where the symbol $\oplus$
denotes the corresponding one-loop SM contributions.

On the experiment side, some decay modes among the all considered  $B \to P( l^+l^-, l\nu, \nu \bar{\nu})$ decays
have been measured by CLEO, BaBar and Belle experiments
\cite{babar-83-032007,cleo-99-041802,belle-648-139,babar-1204-3933,babar-82-112002}.
The $B_s \to K( l^+l^-, l\nu, \nu \bar{\nu})$ decays are now under studying and will be measured by
the LHCb and the forthcoming Super-B experiments \cite{D-2012,Buras-2012}.

On the theory side, the considered semileptonic decays strongly depend on the values and the shape
of the $B/B_s \to P$ form factors. At present, there are various approaches
to calculate the $B/B_s \to (\pi, K)$ transition form factors, such as the lattice QCD technique \cite{HPQCD-2006},
the light cone QCD sum rules (LCSRs) \cite{pball-98,pball-05,jhep04-014,prd83-094031},
as well as the perturbative QCD (pQCD) factorization approach \cite{li-65-014007,yang-npb642,yang-epjc23-28,lu-79-014013,huang-71}.
The direct perturbative calculations of the one-gluon exchange diagram for the $B_{(s)}$ meson transition form
factors suffer from the end-point singularities. Because of these end-point singularities, it was
claimed that the $B\to P$ transition form factors is not calculable perturbatively in QCD \cite{huang-ref-17}.

In the pQCD factorization approach \cite{li-papers}, however, a form factor is
generally written as the convolution
of a hard amplitude with initial-state and final-state hadron distribution amplitudes.
In fact, in the endpoint region the parton transverse momenta $k_{\rm T}$ is not negligible.
If the large double logarithmic term $\alpha_s \ln^2(k_{\rm T})$ and large logarithms  $\alpha_s \ln^2(x)$
are resummed to all orders, the relevant Sudakov form factors from both the $k_{\rm T}$ resummation
and the threshold resummation\cite{pap-resum,li-resum}
can cure the endpoint singularity which makes the perturbative calculation of the
hard amplitudes infrared safe, and then the main contribution comes from the perturbative regions.

In Refs.~\cite{li-65-014007,lu-79-014013}, for example, the authors calculated the
$B \to \pi, \rho$ \cite{li-65-014007} and
$B \to S$ form factors \cite{lu-79-014013} at the leading order by employing the pQCD factorization approach and found that
the values of the corresponding form factors coming from the pQCD factorization approach
agree well with those obtained by using other methods.
In a recent paper\cite{li-85074004}, Li, Shen and Wang calculated the next-to-leading order (NLO)
corrections to the $B \to \pi$ transition form factors at leading twist in the $k_{\rm T}$
factorization theorem. They found that
the NLO corrections amount only up to $30\%$ of the form factors at the large recoil region of the pion.

In this paper, based on the assumption of the $SU(3)$ flavor symmetry, we first extend the NLO results about the
$B \to \pi$ form factors as presented in Ref.~\cite{li-85074004} to the cases of $B \to K$ and $B_s \to K$ directly,
and then calculate the $q^2$-dependence of the differential decay rates and the branching ratios of the considered
$B/B_s$ semileptonic decay modes, and furthermore extract $|V_{ub}|$ based on our calculations.

This paper is organized as follows. In Sec.II, we collect the distribution
amplitudes of the $B/B_s$ mesons and the $\pi, K$ mesons being used
in the calculation and give the $k_{\rm T}$-dependent NLO expressions of the corresponding
transition form factors.
In Sec.III, based on the $k_{\rm T}$ factorization formulism, we calculate and present the
expressions for the $B/B_s \to \pi, K$ transition form factors in the large recoil regions.
The numerical results and relevant discussions are given in Sec.~IV.
And Sec.~V contains the conclusions and a short summary.

\begin{figure}[tbp]
\vspace{-6cm}
\centerline{\epsfxsize=18cm \epsffile{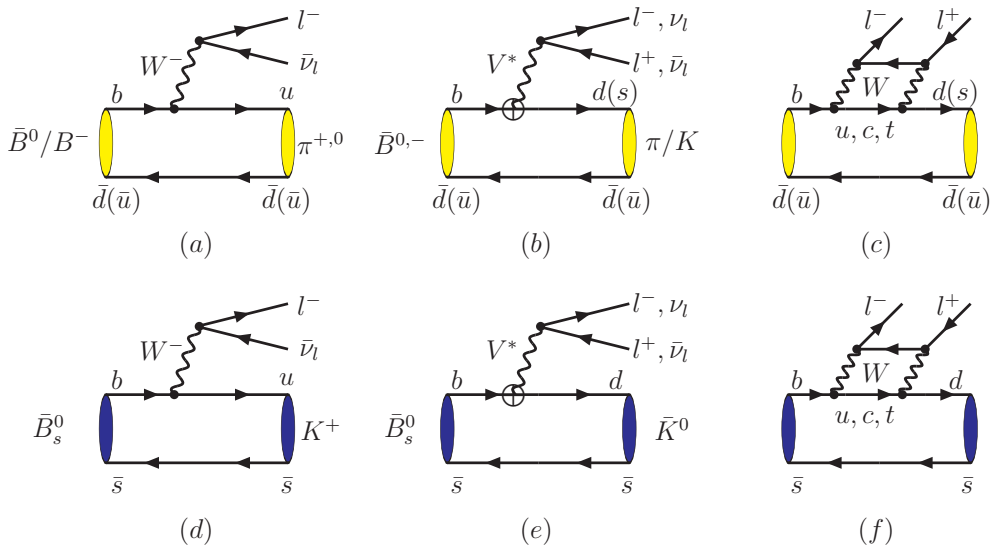} }
\vspace{-14cm}
\caption{ The typical Feynman diagrams for the semileptonic decays
$B/B_s \to (\pi,K)(l^+l^-, l\bar{\nu}, \nu\bar{\nu})$,
where the symbol $\oplus$ in (b) and (e) denotes the flavor-changing neutral current
vertex with $V=\gamma$ and/or $Z$ boson.}
\label{fig:fig1}
\end{figure}


\section{The theoretical framework and NLO corrections} \label{sec:2}

For the sake of simplicity, we use $B$ denotes both the $B$ and $B_s$ meson and
$P$ denotes final meson $\pi$ or $K$ from now on.
As usual, we treat $B$ meson as a heavy-light system.
In the $B$ meson rest frame, with the $m_B$
stands for the mass of $B$ meson, we define the $B$ meson momentum $p_1$
and the final meson $P$ (say $\pi$ or $K$ meson) momentum $p_2$ in the light-cone
coordinates:
\beq
p_1=\frac{m_B}{\sqrt{2}}(1,1,0_{\rm T}),\quad p_2=\frac{m_B}{\sqrt{2}}\eta(0,1,0_{\rm T}),
\eeq
with the energy fraction $\eta=1-q^2/m_B^2$ carried by the final meson (here $q=p_1-p_2$).
The light spectator momenta $k_1$ in the $B$ meson and $k_2$ in the final meson are
parameterized as
\beq
\label{eq:k1k2}
k_1 =(x_1\frac{m_B}{\sqrt{2}},0,k_{1{\rm T}}), \quad
k_2=(0,x_2\eta\frac{m_B}{\sqrt{2}},k_{2{\rm T}}).
\eeq
Because of the final pseudoscalar meson moving along the
minus direction with large momentum, the plus component of its parton’s
momentum should be very small, so it's dropped in the expression of $k_2$.
But the four components of $k_1$ should be of the same order, i.e.
$O(\bar\Lambda)$, with $\bar\Lambda\equiv m_B-m_b$, $m_b$ being the $b$
quark mass. However, since $k_2$ is mainly in the minus direction with
$k_2^-\sim O(m_B)$, the hard amplitudes will not depend on the minus
component $k_1$ as explained below. This is the reason why we do not give
$k_1^-$ in Eq.(\ref{eq:k1k2}) explicitly.

In Ref.~\cite{li-85074004}, the authors derived the $k_{\rm T}$-dependent NLO hard kernel
for the $B \to \pi$ transition form factor. We here use their results directly for $B \to \pi$ transition processes, and
extend the expressions of $B \to \pi$ form factors to the ones for both $B \to K$ and $B_s\to K$ transitions,
under the assumption of $SU(3)$ flavor symmetry. As given in Eq.(56) of Ref.~\cite{li-85074004}, the NLO
hard kernel $H^{(1)}$ can be written as
\beq
H^{(1)}&=& F(x_1,x_2,\eta,\mu_f,\mu,\zeta_1) H^{(0)}\non
&=& \frac{\alpha_s (\mu_f) C_F}{4 \pi} \bigg [  \frac{21}{4}\ln \frac{\mu^2}{m_B^2}
-\left ( \ln \frac{m_B^2}{\zeta_1^2}   + \frac{13}{2}\right) \ln \frac{\mu_f^2}{m_B^2} + \frac{7}{16} \ln^2( x_1 x_2)\non
&&+ \frac{1}{8}\ln^2 x_1  + \frac{1}{4} \ln x_1\ln x_2
+  \left ( 2 \ln \frac{m_B^2}{\zeta_1^2} + \frac{7}{8} \ln \eta - \frac{1}{4} \right) \ln x_1 \non
&&+  \left ( \frac{7}{8} \ln \eta - \frac{3}{2} \right) \ln x_2
+ \left ( \frac{15}{4} - \frac{7}{16} \ln \eta \right ) \ln \eta  \non
&& -\frac{1}{2} \ln \frac{m_B^2}{\zeta_1^2 } \left (  3 \ln \frac{m_B^2}{\zeta_1^2}  + 2 \right )
+ \frac{101}{48} \pi^2 + \frac{219}{16} \bigg ] H^{(0)}.\label{eq:h1h0}
\eeq
where $\zeta_1=25 m_B$\cite{li-85074004}, $\mu_f$ is the factorization scale and set
to the hard scales $t_1$ or $t_2$ as defined in the Appendix, $\eta=1-(p_1-p_2)^2/m_B^2$ is the energy fraction
carried by the final meson, and finally the renormalization scale $\mu$ is defined as \cite{li-85074004}
\beq
t_s (\mu_{\rm f})  = \left \{{\rm Exp} \left[ c_1 + \left(\ln \frac{m_B^2}{\zeta_1^2}
+ \frac{5}{4} \right)  \ln \frac{\mu_{\rm f}^2}{m_B^2 } \right ]  \, x_1^{c_2 } \, x_2^{c_3} \right
\}^{2/21} \, \mu_{\rm f}, \label{ts function}
\eeq
with the coefficients
\beq
c_1 &=& - \left ( \frac{15}{4} - \frac{7}{16} \ln \eta \right ) \ln \eta
+ \frac{1}{2} \ln \frac{ m_B^2}{\zeta_1^2 }   \left ( 3 \ln \frac{ m_B^2}{\zeta_1^2 } + 2 \right )
- \frac{101}{48} \pi^2 - \frac{219}{16} \,,  \non
c_2 &=& - \left ( 2 \ln \frac{ m_B^2}{\zeta_1^2 }  + \frac{7}{8}
\ln \eta - \frac{1}{4} \right )  \, , \non
c_3 &=& -\frac{7}{8} \ln \eta + \frac{3}{2}.
\eeq

In this paper, we use the same distribution amplitudes for $B/B_s$ meson and for the $\pi$ and $K$ meson as those used
in Refs.~\cite{li-85074004,xiao-85094003,pball-05,pball-0605004}.
\beq
\phi_{B}(x,b)&=& N_Bx^2(1-x)^2
\exp\left[-\frac{1}{2}\left(\frac{xm_B}{\omega_b}\right)^2
-\frac{\omega_b^2 b^2}{2}\right] \;,
\eeq
and
\beq
\phi_{B_s}(x,b)&=& N_{B_s} x^2(1-x)^2
\exp\left[-\frac{1}{2}\left(\frac{xm_{B_s}}{\omega_{B_s}}\right)^2
-\frac{\omega_{B_s}^2 b^2}{2}\right] \;,
\eeq
where the normalization factors $N_{B_{(s)}}$ are related to the decay constants $f_{B_{(s)}}$ through
\beq
\int_0^1 dx \phi_{B_{(s)}}(x, b=0) &=& \frac{f_{B_{(s)}}}{2 \sqrt{6}}\;.
\eeq
Here the shape parameter $\omega_b$ has been fixed at $0.40$~GeV by using the rich experimental
data on the $B$ mesons with $f_{B}= 0.21$~GeV. Correspondingly, the normalization constant $N_B$
is $101.4$. For $B_s$ meson, we adopt the shape parameter
$\omega_{B_s} = 0.50$~GeV with $f_{B_s} = 0.23$~GeV, then the corresponding normalization
constant is $N_{B_s} = 63.67$. In order to analyze the uncertainties of
theoretical predictions induced
by the inputs, we can vary the shape parameters $\omega_{b}$ and $\omega_{B_s}$ by
10\%, i.e., $\omega_b = 0.40 \pm 0.04$~GeV and $\omega_{B_s} = 0.50 \pm 0.05$~GeV, respectively.

For the $\pi$ and $K$ mesons, we  adopt the same set of
distribution amplitudes $\phi_i^A(x)$ (the leading twist-2 ) and
$\phi_i^{P,T}(x)$ with $i=(\pi,K)$ as defined in
Refs.~\cite{pball-05,pball-0605004,pball-pi}):
\beq
\phi_i^A(x) &=& \frac{3 f_i}{\sqrt{6} }\, x(1-x) \left [ 1 + a_1
C_1^{3/2}(t) + a_2 C_2^{3/2}(t)+a_4 C_4^{3/2}(t)\right] \;,
\label{eq:phipik-a}\\
\phi^P_i(x) &=& \frac{f_i}{2\sqrt{6}}\, \left [ 1 +\left(30\eta_3
-\frac{5}{2}\rho_i^2\right) C_2^{1/2}(t)  -\, 3\left\{
\eta_3\omega_3 + \frac{9}{20}\rho_i^2(1+6a_2 ) \right\} C_4^{1/2}(t)
\right ]\;, \label{eq:phipik-p}
\\
\phi^\sigma_i(x) &=& \frac{f_i}{2\sqrt{6}}\, x(1-x) \left [ 1
+ \left(5\eta_3 -\frac{1}{2}\eta_3\omega_3 -
\frac{7}{20} \rho_i^2 - \frac{3}{5}\rho_i^2 a_2 \right)C_2^{3/2}(t) \right ]\;,
\label{eq:phipik-t}
\eeq
where $t=2x-1$, $\rho_{\pi(K)}=m_{\pi(K)}/m_0^{\pi(K)}$ are the mass ratios ( here
$m_0^\pi=m_\pi^2/(m_u+m_d)=1.4\pm 0.1$ GeV and $m_0^K=m_K^2/(m_s+m_d)=1.6\pm 0.1$ GeV are the chiral mass
of pion and kaon),
$a_i^{\pi,K}$  are the Gegenbauer moments, while $C_n^{\nu}(t)$ are the
Gegenbauer polynomials
\beq
C_1^{3/2}(t)\, &=&  3\, t \;, \quad
C_2^{1/2}(t)= \frac{1}{2} \left(3\, t^2-1\right), \quad
C_2^{3/2}(t)\, =\, \frac{3}{2} \left(5\, t^2-1\right), \non
C_4^{1/2}(t)\, &=& \, \frac{1}{8} \left(3-30\, t^2+35\, t^4\right), \quad
C_4^{3/2}(t) \,=\, \frac{15}{8} \left(1-14\, t^2+21\, t^4\right) \;.
\label{eq:cii}
\eeq
The Gegenbauer moments appeared in Eqs.~(\ref{eq:phipik-a}-\ref{eq:phipik-t}) are the following
\cite{pball-05,pball-0605004}
\beq
a_1^\pi&=& 0, \quad a_1^K=0.06\pm 0.03, \quad a_2^{\pi,K}=0.25 \pm 0.15, \non
a_4^\pi &=& -0.015, \quad \eta_3^{\pi, K}=0.015, \quad \omega_3^{\pi,K} =-3.
\label{eq:gms}
\eeq


\section{Form factors and  semileptonic decays}\label{sec:3}

\subsection{$B_{(s)}\to\pi, K$ form factors}

The form factors for $B\to P$ transition are defined by \cite{npb592-3}
\beq
\langle P(p_2)|\bar{b}(0)\gamma_{\mu}q(0)|B(p_1)\rangle &=&
\left [(p_1+p_2)_{\mu}-\frac{m_B^2-m_P^2}{q^2}q_{\mu} \right ] F_+(q^2)\non
&& +\frac{m_B^2-m_P^2}{q^2}q_{\mu}F_0(q^2),
\eeq
where $q=p_1-p_2$ is the momentum transfer to the lepton pairs.
In order to cancel the poles at $q^2=0$, $F_+(0)$ should be equal
to $F_0(0)$. For the sake of the calculation, it is convenient to
define the auxiliary form factors $f_1(q^2)$ and $f_2(q^2)$
\beq
\langle P(p_2)|\bar{b}(0)\gamma_{\mu}q(0)|B(p_1)\rangle=
f_1(q^2)p_{1\mu}+f_2(q^2)p_{2\mu}
\eeq
in terms of $f_1(q^2)$ and $f_2(q^2)$, the form factors $F_+(q^2)$ and $F_0(q^2)$ are defined as
\beq
\label{eq:fpfz}
F_+(q^2)&=&\frac12[f_1(q^2)+f_2(q^2)],  \non
F_0(q^2)&=&\frac12 f_1(q^2) \left [1+\frac{q^2}{m_B^2-m_P^2} \right ]
+\frac12 f_2(q^2)\left [1-\frac{q^2}{m_B^2-m_P^2} \right ].
\eeq
As for the tensor operator, there's only one independent
form factor, which is important for the semi-leptonic decay
\beq
\langle P(p_2)|\bar{b}(0)\sigma_{\mu\nu}q(0)|B(p_1)\rangle&=&
i[p_{2\mu}q_\nu-q_{\mu}p_{2\nu}]\frac{2F_\rmt(q^2)}{m_B+m_P},\non
\langle P(p_2)|\bar{b}(0)\sigma_{\mu\nu}\gamma_5q(0)|B(p_1) \rangle
&=&\epsilon_{\mu\nu\alpha\beta}p_2^\alpha q^\beta \frac{2F_\rmt(q^2)}{m_B+m_P}.
\label{eq:ft01}
\eeq
The above form factors are dominated by a single gluon exchange in the lowest
order and in the large recoil regions. The factorization formula for the $B\to P$ form factors
is written as \cite{li-65-014007,yang-epjc23-28}
\beq
\langle P(p_2)|\; \bar{b}(0)\gamma_\mu q(0)|B(p_1)\rangle &=& g_s^2 C_F N_c \int dx_1 dx_2 d^2k_{1{\rm T}}d^2k_{2{\rm T}}
\frac{dz^+d^2 z_{\rm T}}{(2\pi)^3}\frac{dy^-d^2 y_{\rm T}}{(2\pi)^3 } \non
&& \hspace{-2cm}\times e^{-ik_2\cdot y}\langle P(p_2)|\bar q_\gamma^\prime(y)
q_\beta(0)|0\rangle e^{i k_1\cdot z}\langle 0|\bar{b}_\alpha(0)
q_\delta^\prime(z)|B(p_1)\rangle T_{H\mu}^{\gamma\beta;\alpha\delta}.
\eeq

In the hard-scattering kernel, the transverse momentum
in the denominators are retained to regulate the endpoint singularity.
The masses of the light quarks and the mass difference
$(\bar\Lambda)$ between the $b$ quark and the $B$ meson are
neglected. The terms proportional to $k_{1{\rm T}}^2, k_{2{\rm T}}^2$ in the
numerator are dropped, because they are power suppressed compared
to other terms. In the transverse configuration b-space and by
including the Sudakov form factors and the threshold resummation
effects, we obtain the $B\to P$ form factors as following,


\beq
f_1(q^2)&=&16\pi C_Fm_B^2\int dx_1 dx_2\int b_1 db_1 b_2 db_2 \psi_B(x_1,b_1)\non
&& \times \Bigl \{  \left [ r_0 \left ( \phi^p(x_2)-\phi^t(x_2) \right ) \cdot
h_1(x_1,x_2,b_1,b_2) - r_0 x_1 \eta m_B^2\phi^{\sigma}(x_2)
\cdot h_2(x_1,x_2,b_1,b_2) \right ]\non
&& \hspace{1cm} \cdot \alpha_s(t_1) \exp \left [-S_{ab}(t_1) \right ] \non
&& + \left [ x_1 \left ( \eta\phi^a(x_2)-2r_0\phi^p(x_2) \right )+
4 r_0 x_1\phi^p(x_2) \right ] \cdot h_1(x_2,x_1,b_2,b_1) \non
&&
\hspace{1cm} \cdot \alpha_s (t_2)\exp\left [-S_{ab}(t_2) \right] \Bigr \},
\label{eq:f1q2}
\eeq
\beq
f_2(q^2)&=& 16\pi C_Fm_B^2\int dx_1 dx_2\int b_1 db_1 b_2 db_2\psi_B(x_1,b_1)\non
&& \times \Bigl \{ \left [ \left [ \left (x_2\eta+1 \right )\phi^a(x_2)
+2r_0 \left ( \left (\frac{1}{\eta}-x_2 \right )\phi^t(x_2)-x_2\phi^p(x_2)
+3\phi^\sigma(x_2) \right ) \right ]  \right.\non
&& \left.
\cdot h_1(x_1,x_2,b_1,b_2)-r_0 x_1 m_B^2 \left ( 1+x_2\eta \right)
\phi^\sigma(x_2) \cdot h_2(x_1,x_2,b_1,b_2) \right ]
\cdot \alpha_s(t_1)\exp \left [-S_{ab}(t_1) \right ] \non
&&
+ 2 r_0 \left( \frac{x_1}{\eta}+1 \right )
\phi^p(x_2) \cdot h_1(x_2,x_1,b_2,b_1)\cdot \alpha_s(t_2)
\exp\left [-S_{ab}(t_2) \right ] \Bigr \}, \label{eq:f2q2}
\eeq
\beq
F_\rmt(q^2)&=&8\pi C_Fm_B^2\int dx_1 dx_2\int b_1 db_1 b_2 db_2
(1+r_P)\psi_B(x_1,b_1)\non
&& \times \Bigl \{ \left [\ \  r_0 x_1 m_B^2 \phi^\sigma(x_2) \cdot
h_2(x_1,x_2,b_1,b_2)
\right.\non
& & \left.
+ \left [ \phi^a(x_2)-r_0x_2\phi^p(x_2)
+r_0 \left ( \frac{2}{\eta}+x_2 \right )\phi^t(x_2)
+r_0\phi^\sigma(x_2) \right] \cdot h_1(x_1,x_2,b_1,b_2) \right ] \non
&& \hspace{1cm}\cdot \alpha_s(t_1) \exp\left [-S_{ab}(t_1) \right] \non
&& + 2r_0\phi^p(x_2)\left (1+\frac{x_1}{\eta} \right ) \cdot h_1(x_2,x_1,b_2,b_1)
\cdot \alpha_s(t_2)\exp\left [-S_{ab}(t_2) \right] \Bigr \},
\label{eq:ftq2}
\eeq
where $C_F=4/3$ is a color factor, $r_0=m_0^P/m_B=m_P^2/[m_B(m_q+m_{q^\prime})]$,
$r_P=m_P/m_B$ and $m_P$ is the mass of the final pseudoscalar meson,
$m_q$ and $m_q^\prime$ is the mass of the quarks involved in the final meson.
The functions $h_1$ and $h_2$, the scales $t_1$,  $t_2$ and the Sudakov
factors $S_{ab}$ are given in the Appendix A of this paper.
One should note that $f_1(q^2),f_2(q^2)$ and $F_\rmt(q^2)$ as given in
Eqs.~(\ref{eq:f1q2}-\ref{eq:ftq2}) do not including  the NLO correction.
In order to include the NLO corrections, the $\alpha_s$
in Eqs.~(\ref{eq:f1q2}-\ref{eq:ftq2}) should be changed into
$\alpha_s \cdot F(x_1,x_2,\eta,\mu_f,\mu,\zeta_1)$, where the NLO factor
$F(x_1,x_2,\eta,\mu_f,\mu,\zeta_1)$ has been defined in Eq.~(\ref{eq:h1h0}).

\subsection{Semileptonic $B$ and  $B_s$ meson decays }

For the charged current $B \to \pi l^- \bar{\nu}_l$ and $\bar{B}^0_s\to K^+ l^- \bar{\nu}_l$
decays, as illustrated in Fig.1(a) and 1(d),
the quark level transitions are the $b\to ul^-\bar{\nu}_l$ transitions with $l^-=(e^-,\nu^-, \tau^-)$,
the effective Hamiltonian  for such transitions is \cite{operators}
\beq
{\cal H}_{eff}(b\to ul\bar \nu_l)=\frac{G_F}{\sqrt{2}}V_{ub}\;
\bar{u} \gamma_{\mu}(1-\gamma_5)b \cdot \bar l\gamma^{\mu}(1-\gamma_5)\nu_l,
\eeq
where $G_F=1.166 37\times10^{-5} GeV^{-2}$ is the Fermi-coupling constant, $V_{ub}$ is one of the Cabbibo-Kobayashi-Maskawa
(CKM) matrix elements.
The corresponding differential decay widths can be written as \cite{lu-79-014013,p-73-115006}
\beq
\frac{d\Gamma(b\to ul\bar \nu_l)}{dq^2}&=&\frac{G_F^2|V_{ub}|^2}{192 \pi^3
 m_B^3}\frac{q^2-m_l^2}{(q^2)^2}\sqrt{\frac{\left (q^2-m_l^2 \right )^2}{q^2}}
 \sqrt{\frac{\left (m_B^2-m_P^2-q^2 \right )^2}{4q^2}-m_P^2}\non
&& \times \Bigl \{ \left (m_l^2+2q^2 \right ) \left [q^2-\left (m_B-m_P \right )^2 \right ]
\left [ q^2-\left (m_B+m_P \right )^2 \right ]F_+^2(q^2) \non
&& + 3m_l^2\left (m_B^2-m_P^2 \right )^2 F_0^2(q^2)\Bigr \},
 \eeq
where $m_l$ is the mass of the lepton.
If the produced lepton is $e^{\pm}$ or $\mu^{\pm}$, the corresponding mass terms could be neglected,
the above expression then becomes
\beq
\frac{d\Gamma(b\to ul\bar \nu_l)}{dq^2}= \frac{G_F^2
  |V_{ub}|^2}{ 192 \pi^3 m_B^3}\,\lambda^{3/2}(q^2) |F_+(q^2)|^2\,,
\label{eq:dfdq2}
\eeq
where $\lambda(q^2) = (m_B^2+m_P^2-q^2)^2 - 4 m_B^2 m_P^2$ is the phase-space factor.

For those flavor changing neutral current one-loop decay modes, such as $B \to P l^-l^+$
with $P=(\pi, K)$ and $\bar{B}^0_s\to \bar{K}^0  l^-l^+ $ decays, as illustrated in Fig.1,
the quark level transitions are the $b\to (s,d) l^-l^+ $ transitions,
the corresponding effective Hamiltonian  for such transitions is
\beq
{\cal H}_{\mbox{eff}}= -\frac{G_F}{\sqrt{2}}V_{tb}V^*_{tq} \sum_{i=1}^{10}C_i(\mu)O_i(\mu),
\eeq
where $q=(d,s)$, $C_i(\mu)$ are the Wilson coefficients and the local operators $O_i(\mu)$ are given
by \cite{operators}
\beq
\label{eq:operators}
O_1&=&(\bar q_{\alpha}c_{\alpha})_{V-A}(\bar  c_{\beta}b_{\beta})_{V-A},\quad
O_2=(\bar  q_{\alpha}c_{\beta})_{V-A}(\bar  c_{\beta}b_{\alpha})_{V-A},\non
O_3&=&(\bar q_{\alpha}b_{\alpha})_{V-A}\sum_{q^\prime}\left (\bar q^\prime_{\beta}q^\prime_\beta \right )_{V-A},\quad
O_4=(\bar q_{\alpha}b_{\beta})_{V-A}\sum_{q^\prime}(\bar q^\prime_{\beta}q^\prime_{\alpha})_{V-A},\non
O_5&=&(\bar q_{\alpha}b_{\alpha})_{V-A}\sum_{q^\prime}\left (\bar q^\prime_{\beta}q^\prime_\beta \right )_{V+A},\quad
O_6=(\bar q_{\alpha}b_{\beta})_{V-A}\sum_{q^\prime}(\bar q^\prime_{\beta}q^\prime_{\alpha})_{V+A},\non
O_7&=&\frac{e m_b}{8\pi^2}\; \bar q\sigma^{\mu\nu}(1+\gamma_5)b\; F_{\mu\nu},\non
O_9&=&\frac{\alpha_{\rm{em}}}{8\pi}\left (\bar l\gamma_{\mu}l \right ) \left [
\bar q\gamma^{\mu}(1-\gamma_5)b \right ],\quad
O_{10}=\frac{\alpha_{\rm{em}}}{8\pi}\left (\bar l\gamma_{\mu}\gamma_5 l \right )
\left [\bar  q\gamma^{\mu}\left (1-\gamma_5 \right )b \right ],
\eeq
where $q=(d,s)$, $q^\prime=(u,d,c,s,b)$.

For the decays with $b\to sl^+l^-$ transition, for example, the decay amplitude can be written as
\cite{operators}
\beq
{\cal A}(b\to sl^+ l^-)&=&\frac{G_F}{2\sqrt{2}}\frac{\alpha_{\rm{em}}}{\pi}V^*_{ts}V_{tb}
\Bigl \{ C_{10} \left [\bar s\gamma_{\mu}(1-\gamma_5)b \right ]
\left [ \bar l\gamma^{\mu}\gamma_5l \right ]\non
&& + C_9^{eff}(\mu) \left [ \bar s \gamma_{\mu}(1-\gamma_5)b \right ] \left [\bar l\gamma^{\mu}l \right ] \non
&&- 2m_bC_7^{eff}(\mu)\left [ \bar s i\sigma_{\mu\nu}\frac{q^{\nu}}{q^2}
\left (1+\gamma_5 \right )b \right] \left [\bar l\gamma^{\mu}l\right ] \Bigr \},
\eeq
where $C_7^{eff}(\mu)$ and $C_9^{eff}(\mu)$ are the effective Wilson coefficients, defined as
\beq
C_7^{\rm{eff}}(\mu)&=&C_7(\mu)+C^{\prime}_{b\to  s\gamma}(\mu),\label{eq:c7eff}\\
C_9^{\rm{eff}}(\mu)&=&C_9(\mu)+Y_{\rm{pert}}(\hat{s})+Y_{\rm{LD}}(\hat{s}).
\label{eq:c9eff}
\eeq
Here the term $Y_{\rm{pert}}$ represents the short distance perturbative contributions and has been given in
Ref.~\cite{pap-ypert}
\beq
Y_{\rm{pert}}(\hat{s})&=&  h(\hat{m_c},\hat{s})C_0-\frac{1}{2}h(1,\hat{s})(4C_3+4 C_4+3C_5+C_6)\non
&&-\frac{1}{2}h(0,\hat{s})(C_3+3  C_4) + \frac{2}{9}(3C_3 + C_4 +3C_5+ C_6),
\label{eq:ypert}
\eeq
with $C_0=C_1+3C_2+3C_3+C_4+3C_5 +C_6$, $\hat{s}=q^2/m_B^2$, $\hat{m}_c=m_c/m_b$ and $\hat{m}_b=m_b/m_B$,
while the functions $h(z,\hat{s})$ and $h(0,\hat{s})$ in above equation are of the form
\beq
h(z,\hat{s})&=&-\frac{8}{9}\ln\frac{m_b}{\mu}-\frac{8}{9}\ln z+\frac{8}{27}+\frac49 x   \non
& &-\frac29 (2+x)\sqrt{|1-x|} \left\{ \begin{array}{ll}
 (\ln|\frac{\sqrt{1-x}+1}{\sqrt{1-x}-1}|-i\pi),  \quad&(x\equiv\frac{4z^2}{\hat s}<1),\\
2\arctan\frac{1}{\sqrt{x-1}},  &(x\equiv\frac{4z^2}{\hat s}>1),\\
\end{array} \right.\\
h(0,\hat{s})&=&\frac{8}{27}-\frac89 \ln\frac{m_b}{\mu}
-\frac49 \ln\hat s+\frac49 i\pi.
\eeq

The term $Y_{\rm{LD}}(\hat{s})$ in Eq.~(\ref{eq:c9eff}) refers to the long-distance contributions from the resonant states
and will be neglected because they could be excluded by experimental analysis\cite{prl-106161801,bbns2009}.
The term $C^{\prime}_{b \to s\gamma}$ in Eq.~(\ref{eq:c7eff}) is the absorptive part of $b\to s\gamma$ and is given
by\cite{c7eff-pr}
\beq
C^{\prime}_{b \to  s\gamma}(\mu)=i\alpha_s\left \{\frac{2}{9}\eta^{14/23}\left [G_I(x_t)-0.1687 \right ]-0.03C_2(\mu)
\right \},
\eeq
with
\beq
G_I(x_t)=\frac{x_t \left( x_t^2-5x_t-2\right )}{8\left (x_t-1 \right )^3}+\frac{3x_t^2 \ln x_t}{4(x_t-1)^4},
\eeq
where $\eta=\alpha_s(m_W)/\alpha_s(\mu)$ and $x_t=m_t^2/m_W^2$.

The differential decay width of $b\to sl^+l^-$ is given by\cite{lu-79-014013,p-81-074001}
\beq\label{eq-b-sll}
\frac{d\Gamma(b\to sl^+l^-)}{dq^2}&=&
\frac{G_F^2 \alpha^2_{em} |V_{tb}|^2|V^*_{ts}|^2 \sqrt{\lambda(q^2)}}{512 m_B^3\pi^5}
 \sqrt{\frac{q^2-4m_l^2}{q^2}}\frac{1}{3q^2} \non
&&\times\bigg [ 6m_l^2 |C_{10}|^2(m_B^2-m_P^2)^2F_0^2(q^2) \non
&&+(q^2+2m_l^2)\lambda(q^2) \bigg | C_9^{eff} \; F_+(q^2)+\frac{2C_7^{eff}\; (m_b-m_s) F_\rmt(q^2)}{m_{B}+m_{P}}\bigg|^2  \non
&&+ |C_{10}|^2(q^2-4m_l^2)\; \lambda(q^2)\; F_+^2(q^2)
\bigg ],
\eeq
where $\alpha_{em}=1/137$ is the fine structure constant. For $b \to d l^+l^-$ decays, it is easy to derive the differential
decay width from above equation by a simple replacement $V_{ts}\to V_{td}$ and $m_s\to m_d$.

Finally, the effective Hamiltonian  for  $b \to s \nu \bar \nu$ transition is
\beq
\label{eq-Hanunu}
H_{b \to s\nu \bar \nu}&=& \frac{G_F}{\sqrt{2}} \frac{\alpha_{em}}{2 \pi \sin^2(\theta_W)} V_{tb} V_{ts}^* \eta_X X(x_t) \;
\left[ {\bar s}\gamma^\mu (1-\gamma_5) b \right ] \left[ {\bar \nu} \gamma_\mu(1-\gamma_5) \nu \right]\non
&=& C_L^{b\to s}\; O_L^{b\to s},
\eeq
where  $\theta_W$ is the Weinberg angle with $\sin^2(\theta_W)=0.231$, the  function $X(x_t)$ can be found in
Ref.~\cite{operators}, while $\eta_X\approx 1$ is the QCD factor \cite{operators}.
The corresponding differential decay width can be written as
\beq
\label{eq-bsvv}
\frac{d\Gamma(b \to s \nu \bar \nu)}{dq^2}=3 \frac{|C_L^{b\to s}|^2\lambda^{3/2}
(m_{B}^2,m_{P}^2,q^2)}{96 m_{B}^3 \pi^3}|F_+(q^2)|^2.
\eeq
The factor 3 in above equation arises form the summation over
the three neutrino generations. For $b\to d\nu\bar\nu$ transition, we can obtain the differential decay widths
easily also by simple replacements: $ |V_{ts}^*| \to |V_{td}^*|$ and $m_s\to m_d$.

\section{Numerical results and discussions} \label{sec:4}

In the numerical calculations we use the following input parameters (the masses and decay constants are
all in unit of GeV) \cite{pdg2010}:
\beq
\Lambda_{\bar{MS}}^{(f=4)}&=&0.287,\quad f_\pi=0.13,\quad f_K=0.16,\quad f_B=0.21, \non
f_{B_s}&=& 0.236,\quad  m_{B^\pm}=5.2792,\quad m_{B^0}=5.2795, \quad m_{B_s^0}=5.3663,\non
\tau_{B^\pm}&=&1.638\; ps,\quad \tau_{B^0}=1.525\; ps, \quad \tau_{B_s^0}=1.472\; ps,\non
m_{\pi^\pm}&=& 0.1396,\quad m_{\pi^0}=0.135, \quad m_{K^\pm}=0.4937,\quad  m_{K^0}=0.4976,\non
m_\tau &=& 1.777, \quad m_b=4.8, \quad m_W=80.4, \quad m_t=172.
\eeq
For relevant CKM matrix elements we use $|V_{tb}|=0.999$, $|V_{ts}|=0.0403^{+0.0011}_{-0.0007}$,
$|V_{td}/V_{ts}|=0.211\pm0.001\pm0.005$ \cite{pdg2010}.

\subsection{Form factors in the pQCD factorization approach}

By using the definitions in Eqs.~(\ref{eq:fpfz},\ref{eq:ft01}) and the expressions in Eqs.(\ref{eq:f1q2}-\ref{eq:ftq2}), we
can calculate the values of the form factors $F_0(q^2)$, $F_+(q^2)$ and $F_\rmt(q^2)$ for given value of $q^2$ in the region
of $0\leq q^2 \leq (M_B-m_P)^2$.
But one should note that the pQCD predictions for the considered form factors are reliable only for small or moderate
values of $q^2$: say $0\leq q^2 \leq 12$ ${\rm GeV}^2$. For the form factors in the larger $q^2$ region, one has to
make an extrapolation for them from the lower $q^2$ region to larger $q^2$ region.

For the form factor $F_0(q^2)$ of $B/B_s\to (\pi, K)$ transition, we make the extrapolation by using
the pole model parametrization
\beq
 F_0(q^2)=\frac{F_0(0)}{1-a(q^2/m_B^2)+b(q^2/m_B^2)^2}, \label{eq:pole-1}
\eeq
where $a, b$ are the constants to be determined by the fitting procedure.
In Table \ref{tab-f0}, we list the LO and NLO pQCD predictions for the form
factors $F_0(0)$ and the corresponding
parametrization constant $a$ and $b$ for $B\to(\pi,K)$ and $B_s \to K$
transitions extracted through  the fitting. The first error
of $(F_0(0), a, b)$ in Table  \ref{tab-f0}
comes from the uncertainty of $\omega_b=0.40\pm 0.04$ GeV or
$\omega_{B_s}=0.50\pm 0.05$ GeV, the second ne is induced by
$a_1^K=0.06\pm 0.03$ and/or  $a_2^{\pi, K}=0.25 \pm 0.15$。The errors from the uncertainties
of $m_0^{\pi,K}$, $V_{ts}$ and $|V_{td}/V_{ts}|$ are very small and have been neglected.

\begin{table}[thb]
\begin{center}
\caption{The pQCD predictions for form factors $F_0(0)$ and the parametrization
constant $a$ and $b$
for $B\to(\pi,K)$ and $B_s \to K$ transitions, at the LO and NLO level
respectively. The two errors come from the uncertainties of $\omega_b$ or $\omega_{B_s}$,
and $a_1^K$ and/or $a_2^{\pi,K}$, respectively.}
\label{tab-f0}
\begin{tabular}{l| l| l| l} \hline\hline
\ \ \ &$ \ \ F_0(0)_{LO}$ &$\ \ a_{LO}$ &$ \ \ b_{LO}$  \\
 \hline
\ \ \ $B\to\pi$\; &$0.22^{+0.03}_{-0.02}\pm 0.01$\;&$0.58 \pm 0.01 \pm 0.03$\;
&$-0.15 \pm 0.01 \pm 0.01$\;\\  \hline
\ \ \
$B\to K$   &$0.27^{+0.04}_{-0.03}\pm 0.01$ &$0.60 \pm 0.01 \pm 0.03$
&$-0.15^{+0.01 +0.01}_{-0.00-0.02}$\\   \hline
 \ \ \
$B_s\to K$\;   &$0.22\pm 0.03  \pm0.01$ &$0.61^\pm 0.01^{+0.04}_{-0.03}$
 &$-0.16\pm 0.00 ^{+0.02}_{-0.01}$\\ \hline\hline
\ \ \ &
$\ \ F_0(0)_{NLO}$ &$\ \ a_{NLO}$ &$\ \ b_{NLO}$ \\  \hline
\ \ \ $B\to\pi$\; &$0.26^{+0.04}_{-0.03} \pm0.02$\;
&$0.50\pm 0.01 ^{+0.05}_{-0.04}$ &$-0.13\pm 0.01 \pm 0.01$\;\\  \hline
\ \ \ $B\to K$   &$0.31 \pm 0.04 \pm 0.02$
&$0.53\pm 0.01 ^{+0.05}_{-0.04}$
&$-0.13\pm 0.01^{+0.02}_{-0.01}$\\   \hline
 \ \ \ $B_s\to K$\;   &$0.26^{+0.04}_{-0.03} \pm0.02$
 &$0.54 \pm 0.00 \pm 0.05$  &$-0.15\pm 0.01 \pm 0.01$\\  \hline\hline
\end{tabular}
\end{center}
\end{table}

\begin{figure}[thb]
\begin{center}
\centerline{\epsfxsize=10cm \epsffile{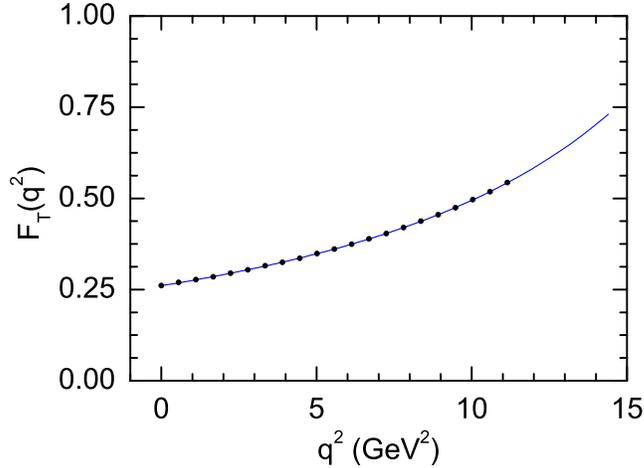}}
\caption{ The pQCD predictions for the form factors $F_{\rm T}(q^2)$ for $B\to \pi$ transition,
where the dots refer to the pQCD predictions for the given points of $q^2$ in the range
of $0\leq q^2 \leq 12 GeV^2$,
while the solid curve stands for the fitted curve  at the NLO level.}
\label{fig:fig2}
\end{center}
\end{figure}

For the form factors $F_+(q^2)$ and $F_\rmt(q^2)$, the pole model parametrization as given in Eq.~(\ref{eq:pole-1})
does not work, and we have to use other proper parametrization method.
In this paper, we use the Ball/Zwicky (BZ) parametrization method
\cite{pball-05,param-bz,param-bz2}.
It includes the essential feature that $F_+(q^2)$ and $F_\rmt(q^2)$ have a pole at $q^2 = m_{B^*}^2$,
with $B^*(1^-)$ is a narrow resonance with $m_{B^*} = 5.325~GeV$ and
$m_{B_s^*} = 5.415~GeV$, which are expected to have a distinctive impact on the form factor.

For the form factors $F_{+}(q^2)$ and $F_\rmt(q^2)$ of $B/B_s\to (\pi, K)$
transition, we make the extrapolation by using the BZ parametrization
\beq
\label{pa-bz}
F_i(q^2) = F_i(0)\left(\frac{1}{1-q^2/m_{B_{(s)}^*}^2} +
\frac{r q^2/m_{B_{(s)}^*}^2}{\left(1-q^2/m_{B_{(s)}^*}^2\right)
\left(1-\alpha\,q^2/m_{B_{(s)}}^2\right)} \right),
\eeq
where $\alpha$ and $r$ are the shape parameters to be determined by the
fitting procedure, the same as for the case of $F_0(q^2)$.
In Table \ref{tab-fp}, we list the pQCD predictions for
the form factors $F_+(0)$, $F_{\rm T}(0)$ and the corresponding shape parameters
$(\alpha, r)$ for $B\to(\pi,K)$ and $B_s \to K$ transitions at the LO and NLO
level. In Fig.~2, we show the pQCD predictions for the form factors $F_{\rm T}(q^2)$ for $B\to \pi$
transition, where the dots refer to the pQCD predictions for each given value of $q^2$ in the large-recoil
range of $0\leq q^2 \leq 12 GeV^2$, while the solid curve stands for the fitted curve  at the NLO level,
obtained through fitting by using the Eq.~(\ref{pa-bz}).

\begin{table}[thb]
\begin{center}
\caption{The same as in Table \ref{tab-f0}, but for the pQCD predictions
for the form factors $F_+(0)$, $F_{\rm T}(0)$ and the corresponding
shape parameters $\alpha$ and $r$ at the LO and the NLO level.
And the two errors comes from the uncertainties of $\omega_b$ or
$\omega_{B_s}$, and $a_1^K$ and/or $a_2^{\pi,K}$, respectively.}
\label{tab-fp}
\begin{tabular}{l| l| l| l}
\hline\hline
 \ \ \ &$ \ \ F_+(0)_{LO}$ &$\ \ \alpha_{LO}$ &$\ \ r_{LO}$ \\
 \hline
 \ \ \ $B\to\pi$\; &$0.22^{+0.03}_{-0.02}\pm0.01$\;
 &$0.61^{+0.00}_{-0.01} \pm 0.01$\;
 &$0.51\pm 0.00\pm 0.03 $\;  \\
  \hline
 \ \ \ $B\to K$   &$0.27^{+0.04}_{-0.03} \pm0.01 $
 &$0.62^{+0.00}_{-0.01} \pm 0.01$
 &$0.58\pm 0.00 \pm 0.03 $  \\
  \hline
 \ \ \ $B_s\to K$\;   &$0.22 \pm 0.03 \pm0.01$
 &$0.64\pm 0.00\pm 0.01$
 &$0.56\pm 0.00 \pm ^{+0.04}_{-0.03}$  \\ \hline\hline
 \ \ \ &$\ \ F_{\rm T}(0)_{LO}$ &$\ \ \alpha_{LO}$ &$\ \ r_{LO}$ \\
 \hline
 \ \ \ $B\to\pi$\; &$0.23 \pm 0.03  \pm 0.01$\;
 &$0.69^{+0.00}_{-0.01} \pm 0.01$\;
 &$0.55 \pm 0.01\pm 0.03$\;   \\   \hline
 \ \ \ $B\to K$   &$0.30^{+0.04}_{-0.03} \pm0.01$
 &$0.71^{+0.00}_{-0.01} \pm 0.01$
 &$0.58 \pm 0.01 \pm 0.03$  \\   \hline
 \ \ \ $B_s\to K$\;   &$0.25^{+0.04}_{-0.03} \pm0.01$
 &$0.71^{+0.01}_{-0.00}\pm 0.01$
 &$0.59\pm 0.00 \pm 0.03$    \\  \hline\hline
 \ \ \ &$\ \ F_+(0)_{NLO}$ &$\ \ \alpha_{NLO}$ &$\ \ r_{NLO}$  \\
 \hline
 \ \ \ $B\to\pi$\; &$0.26^{+0.04}_{-0.03}\pm 0.02$\;
 &$0.52\pm 0.01 \pm 0.03$\;
 &$0.45\pm 0.00 ^{+0.05}_{-0.04}$\;    \\
  \hline
 \ \ \ $B\to K$   &$0.31\pm 0.04 \pm 0.02$
 &$0.54\pm 0.01 ^{+0.02}_{-0.03}$
 &$0.50\pm 0.00 \pm 0.05$    \\   \hline
 \ \ \ $B_s\to K$\; &$0.26^{+0.04}_{-0.03}\pm0.02$
 &$0.57\pm 0.01\pm 0.02 $
 &$0.50\pm 0.01 \pm 0.05$    \\  \hline\hline
 \ \ \ &$\ \ F_{\rm T}(0)_{NLO}$ &$\ \ \alpha_{NLO}$ &$\ \ r_{NLO}$  \\  \hline
 \ \ \ $B\to\pi$\; &$0.26^{+0.04}_{-0.03}\pm 0.02$\;
 &$0.65^{+0.01}_{-0.02} \pm0.01 $\;
 &$0.50\pm 0.00\pm 0.00$\;    \\   \hline
 \ \ \ $B\to K$  &$0.34^{+0.05}_{-0.04}\pm0.02$
 &$0.67\pm 0.01 \pm 0.01$
 &$0.53 \pm0.00 ^{+0.05}_{-0.04}$    \\  \hline
 \ \ \ $B_s\to K$ &$0.28\pm 0.04 \pm 0.02$
 &$0.69\pm 0.01 \pm 0.01 $
 &$0.53^{+0.01 +0.04}_{-0.00-0.02} $    \\  \hline\hline
\end{tabular}
\end{center}
\end{table}

\begin{figure}[thb]
\begin{center}
\centerline{\epsfxsize=5.5cm \epsffile{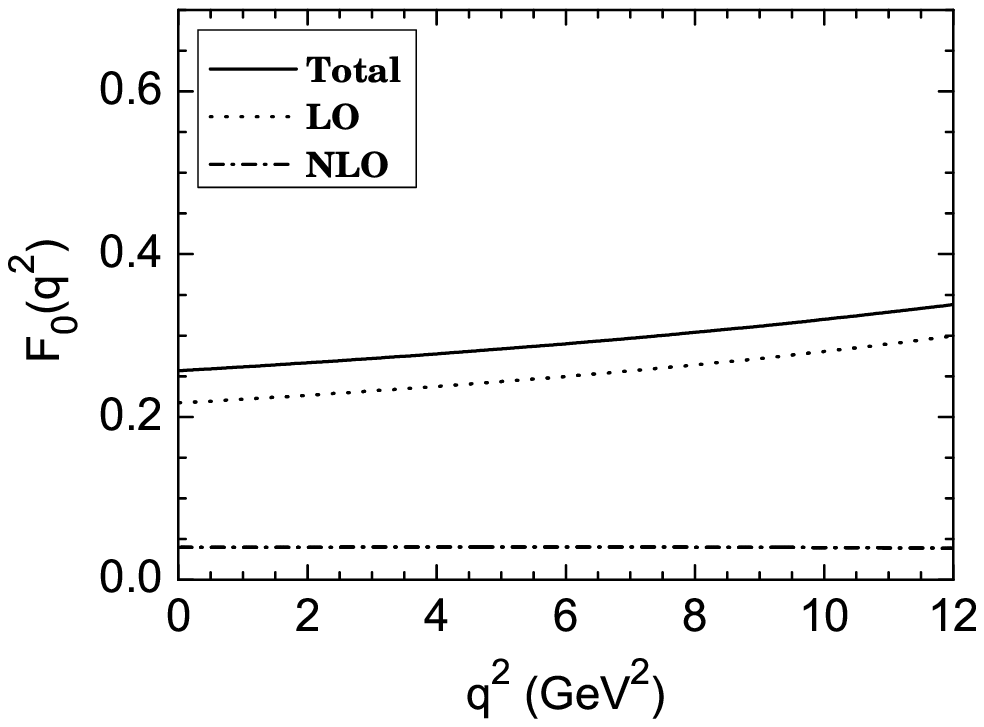}
\epsfxsize=5.5cm \epsffile{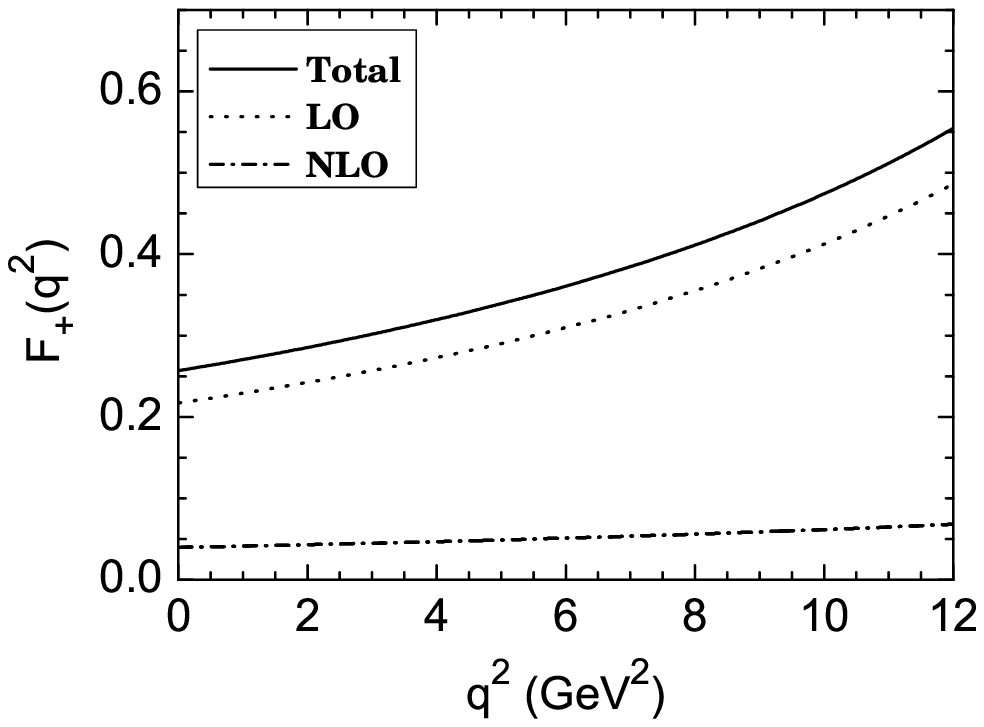}
\epsfxsize=5.5cm\epsffile{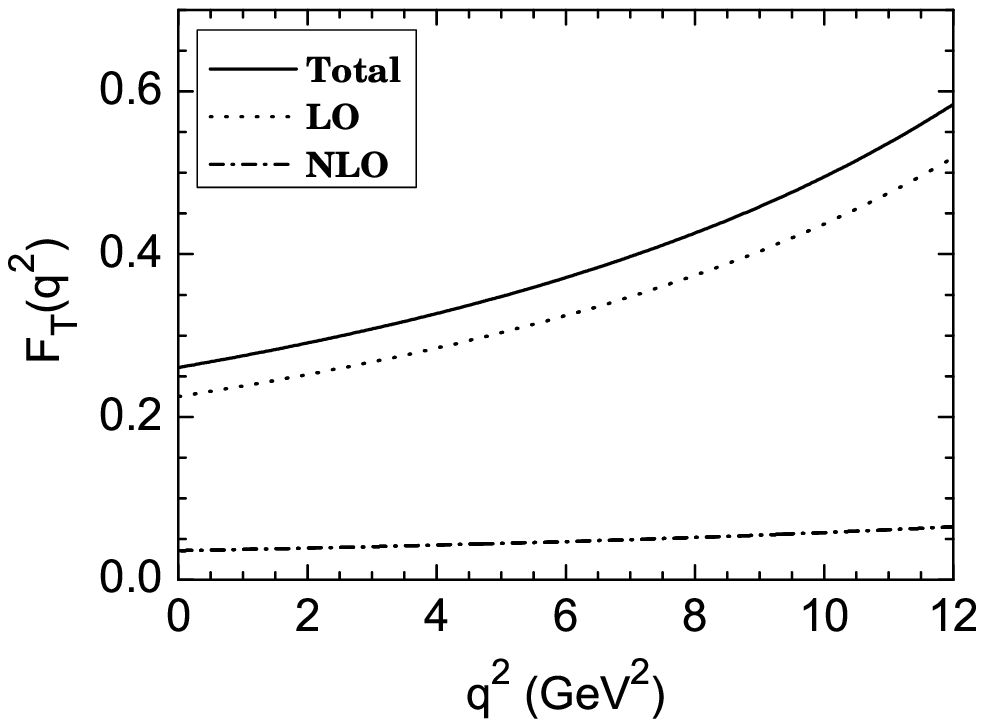}  }
\caption{ The pQCD predictions for the $q^2$-dependence of
$F_{0,+,\rmt}(q^2)$ for $B\to \pi$ transition,
where the dots curve and dot-dashed curve shows the LO and NLO
part respectively, and the solid curve stands for the total value at the NLO level.}
\label{fig:fig3}
\end{center}
\end{figure}

\begin{figure}[thb]
\begin{center}
\centerline{\epsfxsize=5cm \epsffile{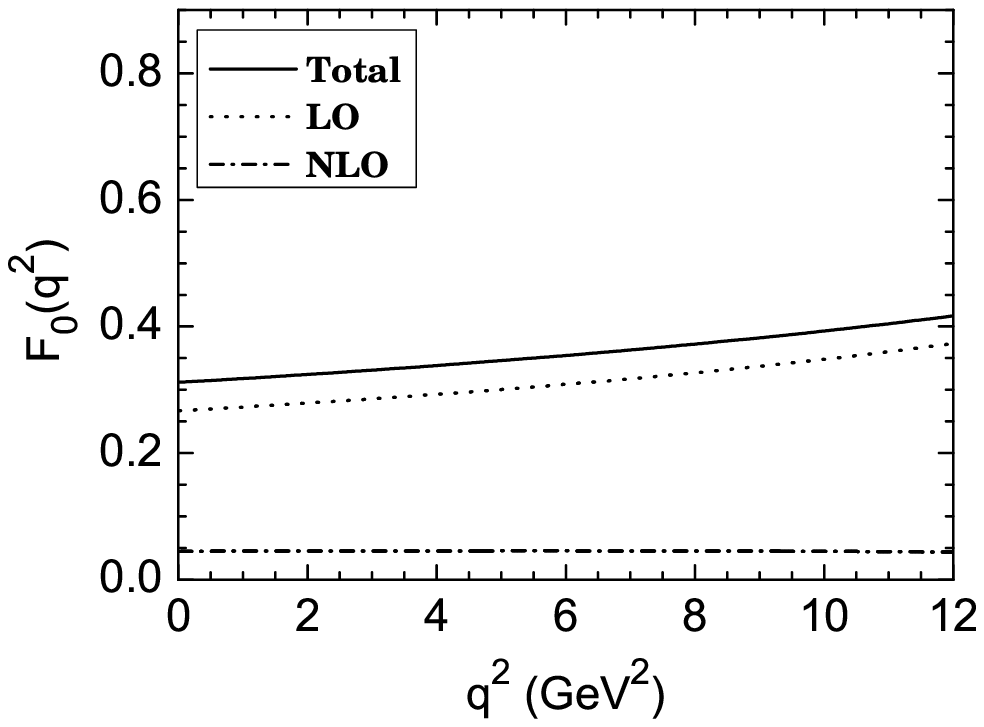}
\epsfxsize=5cm \epsffile{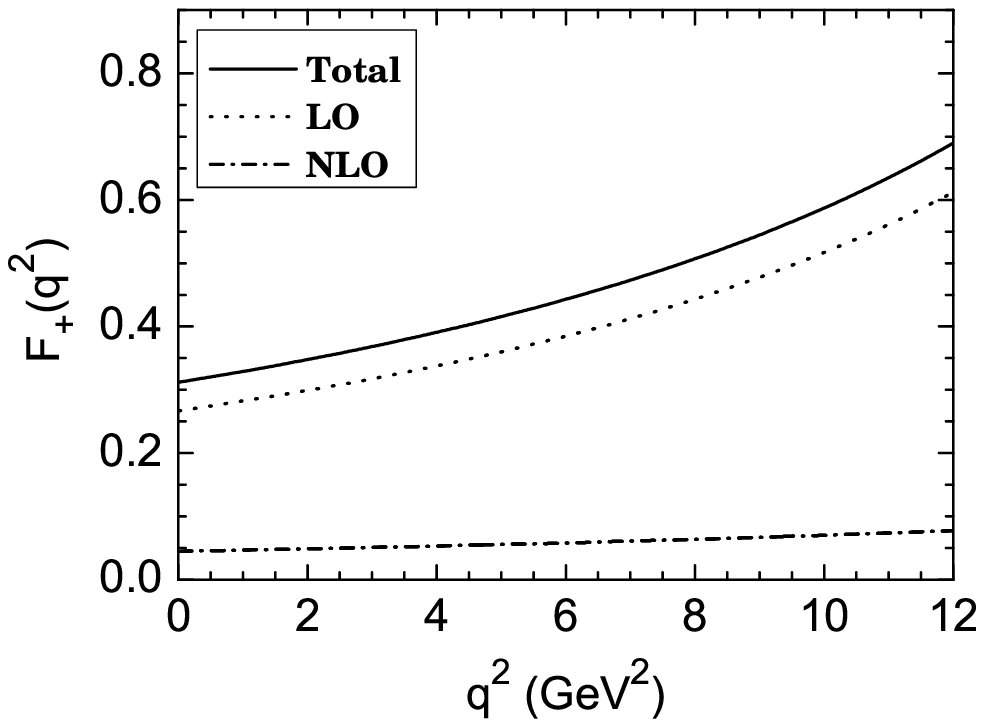}
\epsfxsize=5cm\epsffile{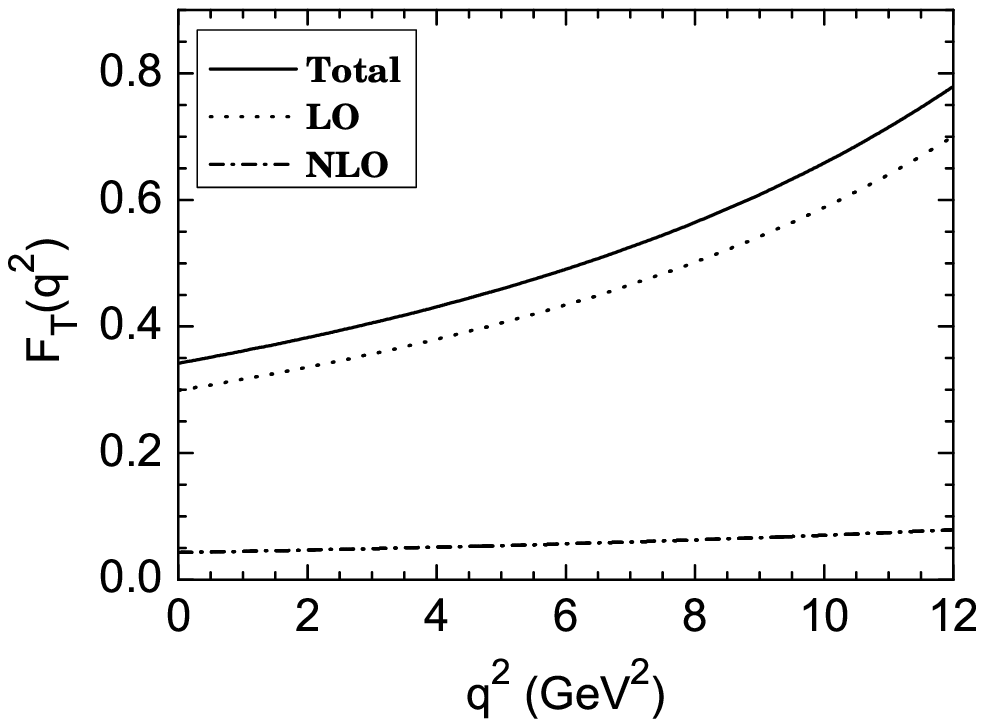}  }
\caption{ The same as in Fig.~\ref{fig:fig3} but for $B \to K $ transition.}
\label{fig:fig4}
\end{center}
\end{figure}

\begin{figure}[thb]
\begin{center}
\centerline{\epsfxsize=5cm \epsffile{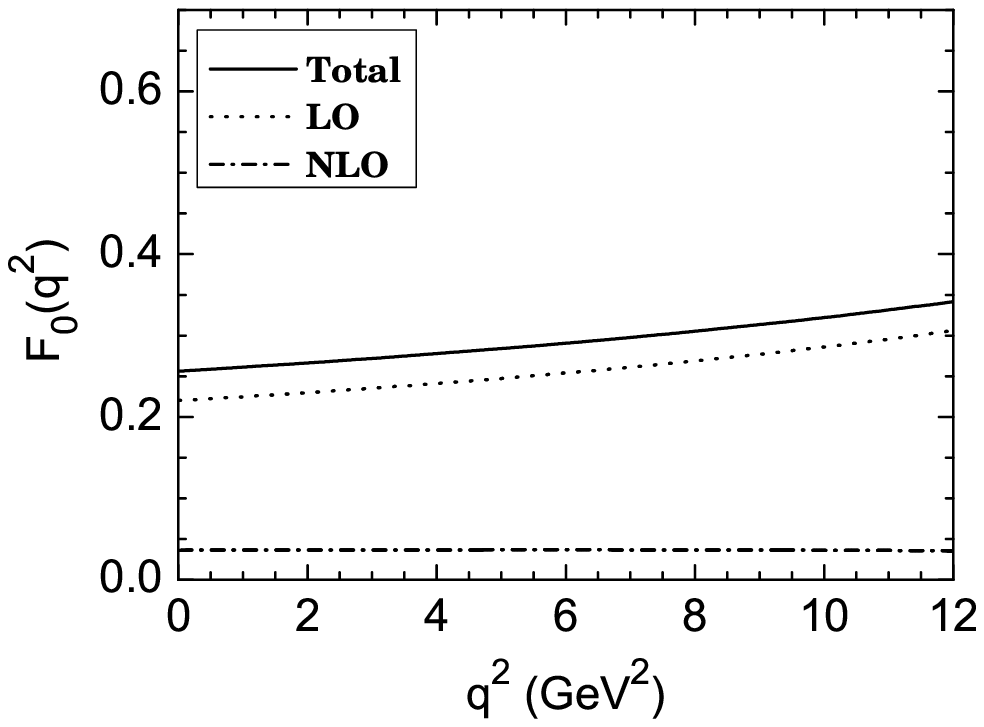}
\epsfxsize=5cm \epsffile{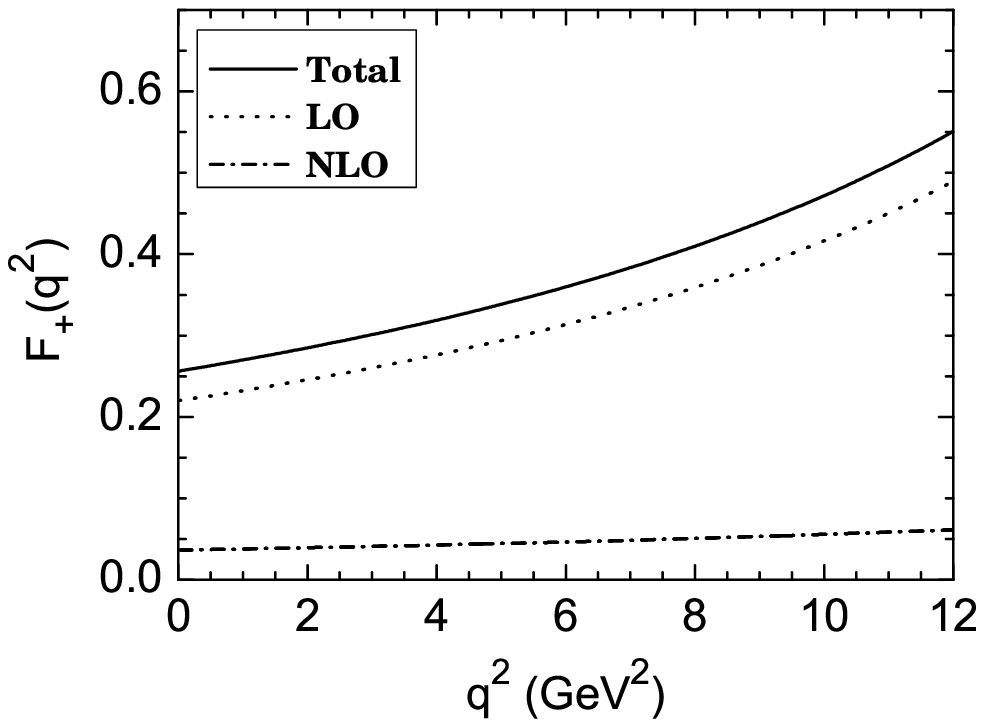}
\epsfxsize=5cm\epsffile{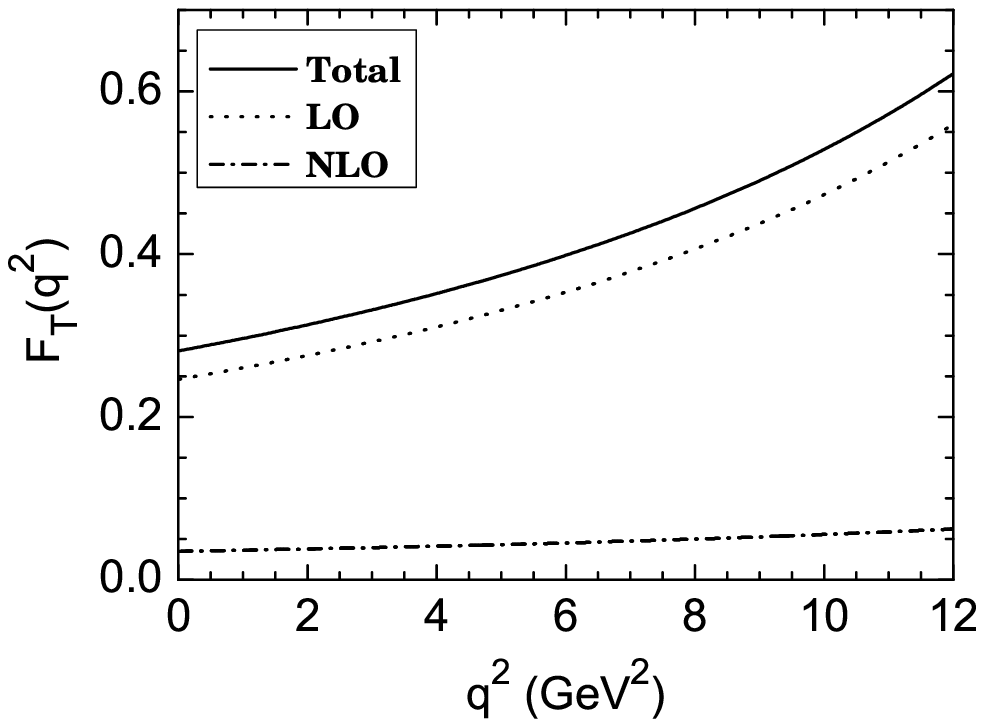}  }
\caption{ The same as in Fig.~\ref{fig:fig3} but for $B_s \to K $ transition.}
\label{fig:fig5}
\end{center}
\end{figure}

In Figs.~(\ref{fig:fig3}-\ref{fig:fig5}), we show the pQCD predictions for the
$q^2$-dependence of the form factors $F_{0,+,\rmt}(q^2)$ at the leading order
(dots curves) and the next-to-leading order (solid curve) for the considered
$B \to (\pi, K)$ and $B_s \to K$ transitions, respectively.

From the numerical results as listed in Table \ref{tab-f0} and \ref{tab-fp}
and the $q^2$-dependence as illustrated in Figs.(\ref{fig:fig3}-\ref{fig:fig5}),
one can see that:
\begin{enumerate}
\item[(i)]
For the considered $B\to (\pi, K)$ and $B_s \to K$ transitions,
the  NLO pQCD predictions for the form factors $F_{0,+,\rmt}(0)$ agree well
with the values estimated from the LCSR or other methods\cite{pball-05}.

\item[(ii)]
$F_{0}(0)$ equals to $F_{+}(0)$ by definition, but they have different $q^2$-dependence
as illustrated by Figs.\ref{fig:fig3}-\ref{fig:fig5}. We also observe the pattern of the relative
strength of the form factors:
\beq
F_{0,+}^{B\to \pi}(0) = F_{0,+}^{B_s\to K}(0) \lesssim F_{0,+}^{B\to K}(0), \\
F_{\rm T}^{B\to \pi}(0) \lesssim F_{\rm T}^{B_s\to K}(0) \lesssim F_{\rm T}^{B\to K}(0),
\label{eq:ff001}
\eeq
which is consistent with the general expectation.

\item[(iii)]
The LO part of the form factors dominate the total contribution, the NLO part
is only around $20\%$. The form factor $F_0(q^2)$ has a relatively weak $q^2$-dependence,
but  $F_{+}(q^2)$ and $F_{\rm T}(q^2)$ show a little stronger $q^2$-dependence
when compared with  $F_0(q^2)$.

\end{enumerate}

\subsection{Decay widths and branching ratios}

By using the relevant formula and the input parameters as defined or given
in previous sections, it is straightforward to calculate the branching
ratios for all the considered decays.

\begin{figure}[thb]
\begin{center}
\vspace{-0.5cm}
\centerline{\epsfxsize=5cm \epsffile{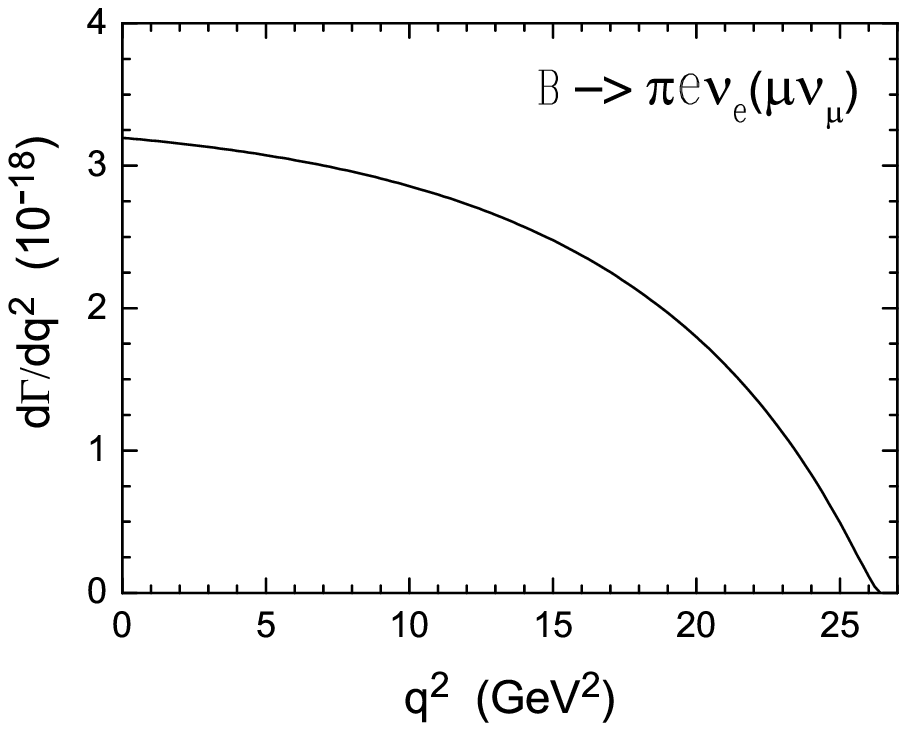}
\epsfxsize=5cm \epsffile{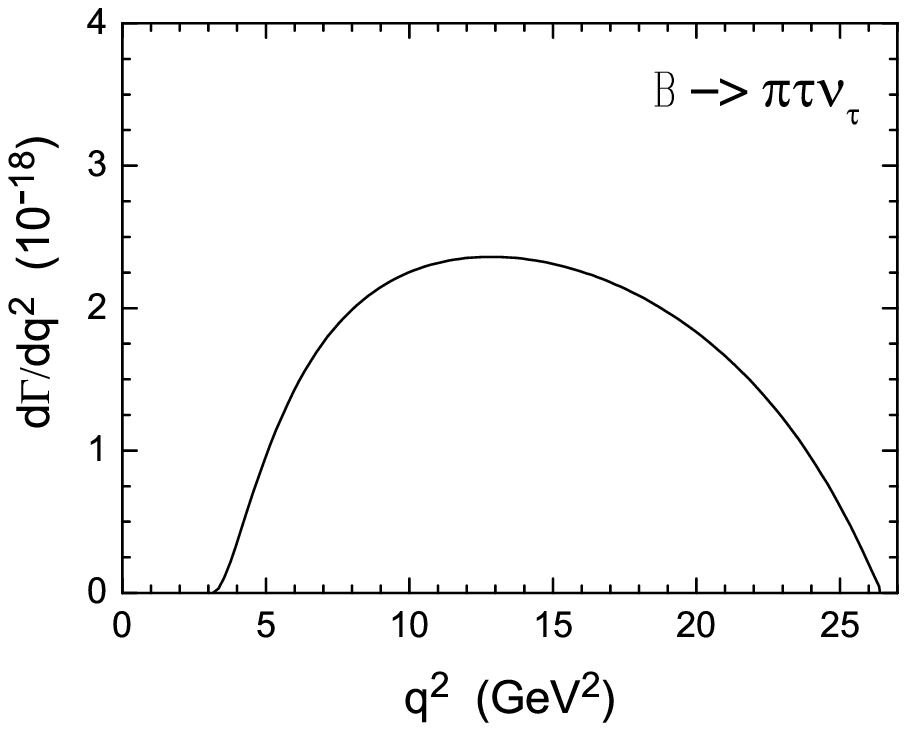}}
\centerline{\epsfxsize=5cm \epsffile{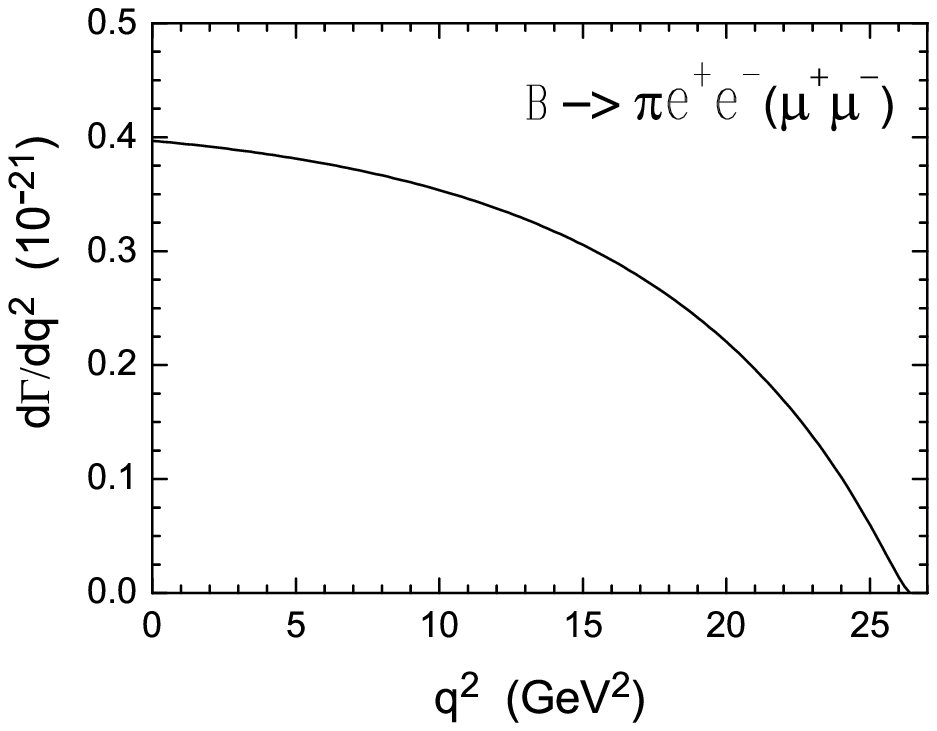}
\epsfxsize=5cm \epsffile{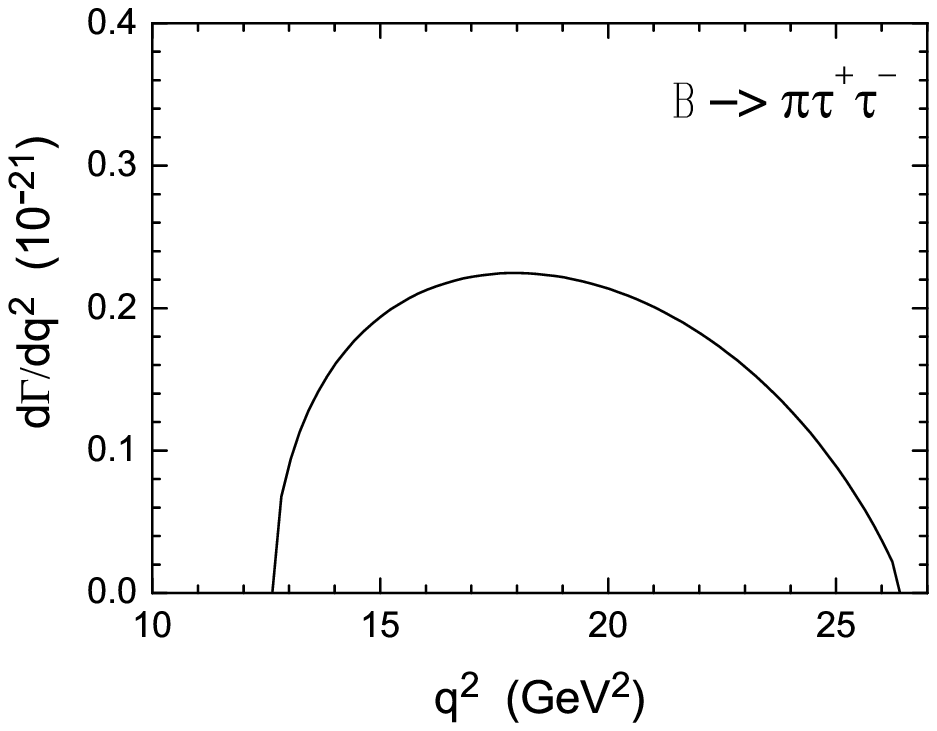}
\epsfxsize=5cm\epsffile{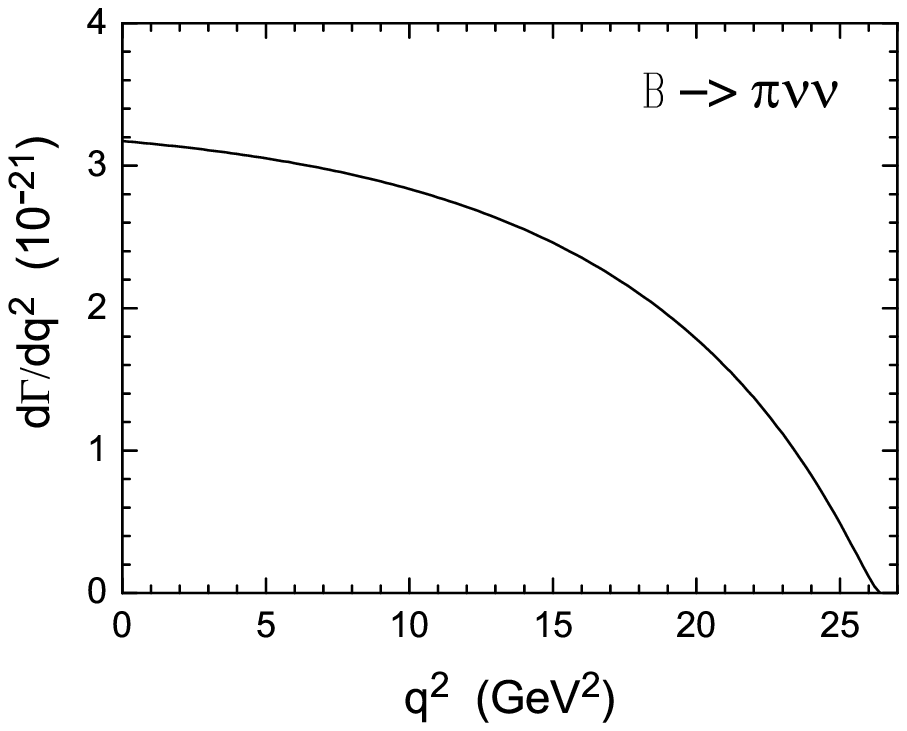}  }
\centerline{\epsfxsize=5cm \epsffile{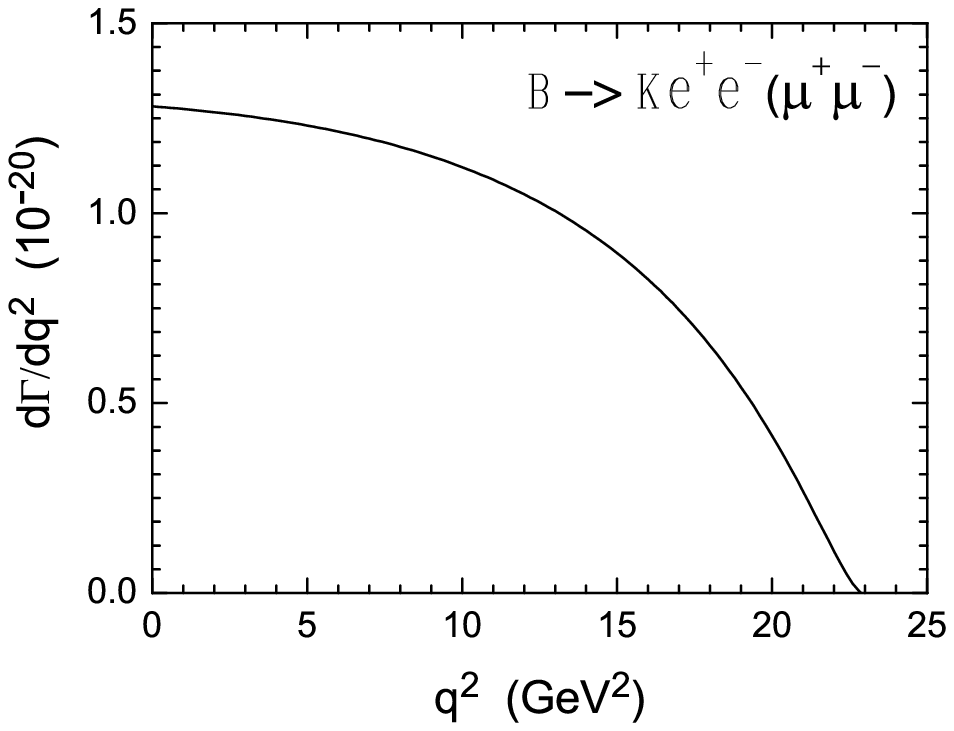}
\epsfxsize=5cm \epsffile{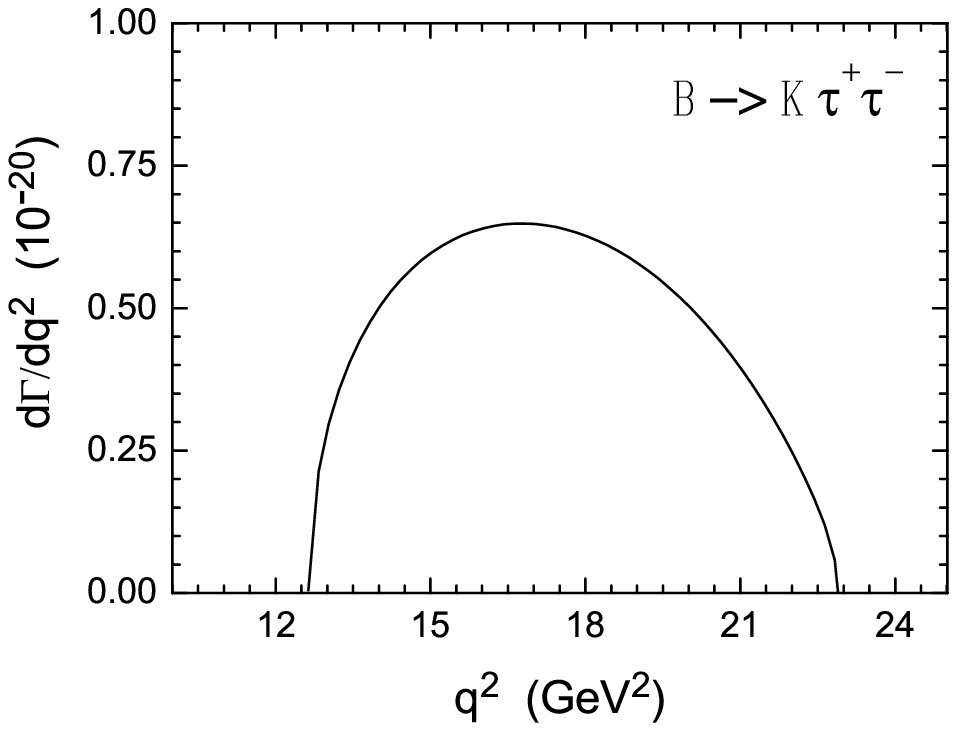}
\epsfxsize=5cm\epsffile{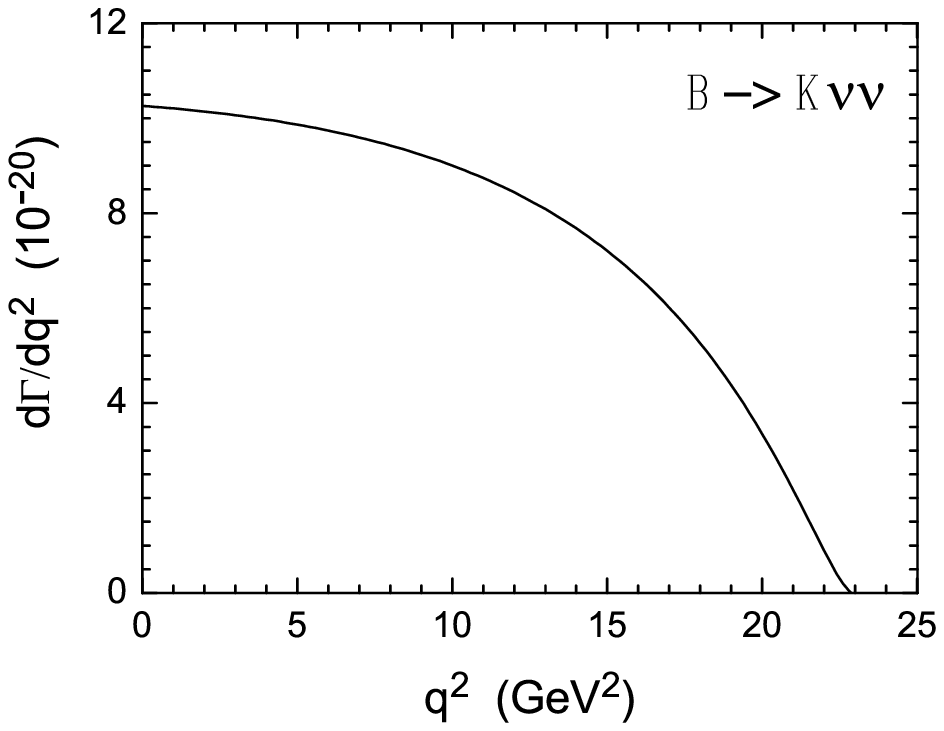}  }
\caption{The $q^2$-dependence of the differential decay rates $d\Gamma/dq^2$
for the decay processes with the $B\to\pi$ and $B \to K$ transitions.}
\label{fig:fig6}
\end{center}
\end{figure}

\begin{figure}
\begin{center}
\vspace{-1cm}
\centerline{\epsfxsize=5cm \epsffile{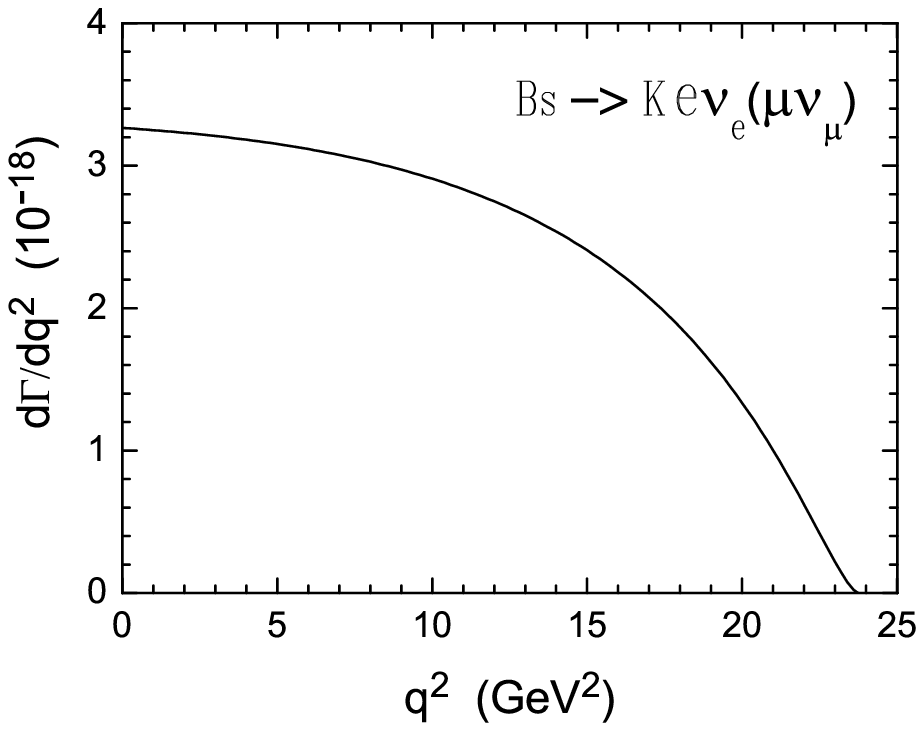}
\epsfxsize=5cm \epsffile{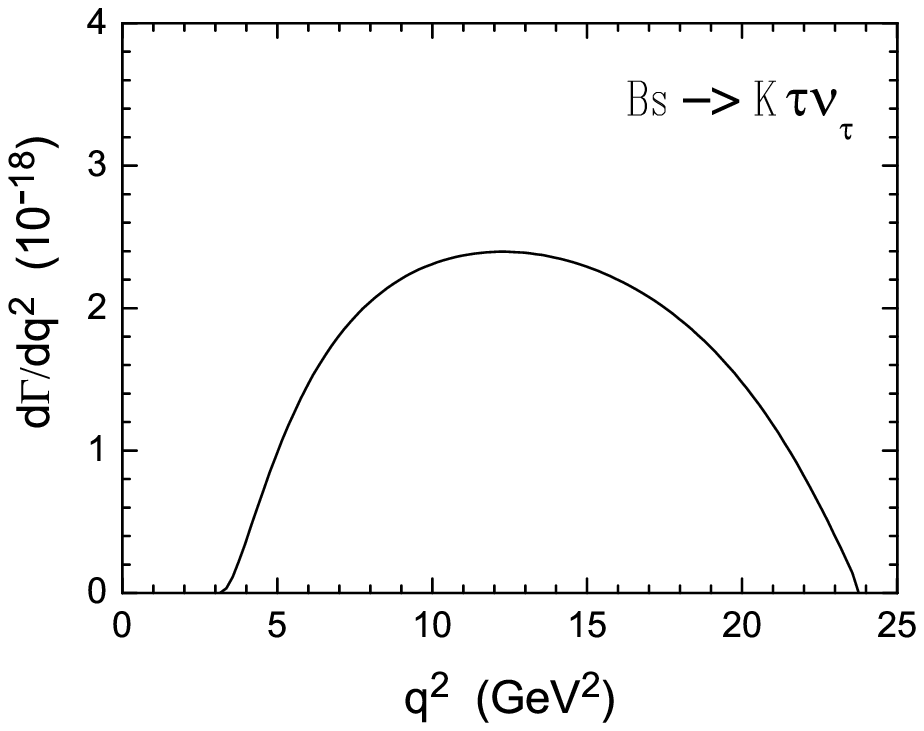}}
\centerline{\epsfxsize=5cm \epsffile{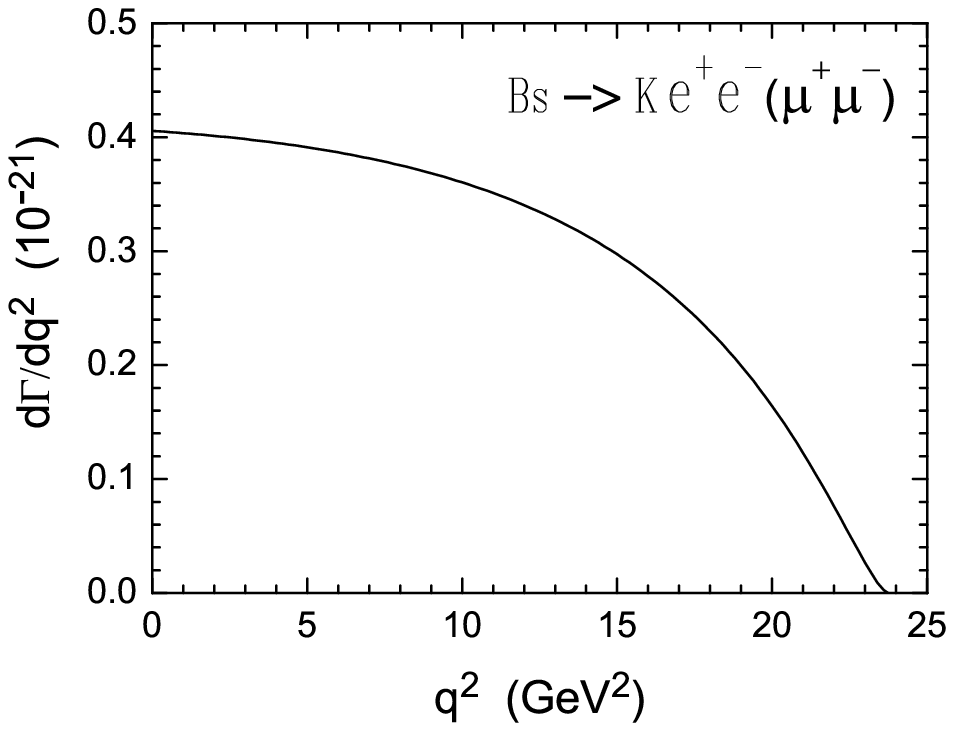}
\epsfxsize=5cm \epsffile{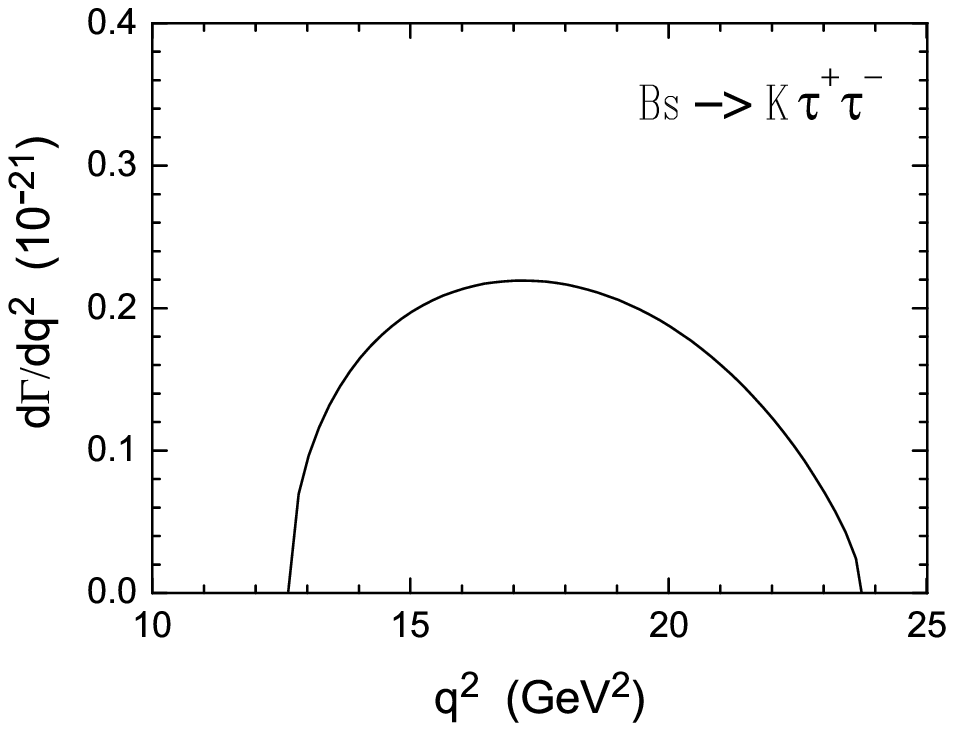}
\epsfxsize=5cm\epsffile{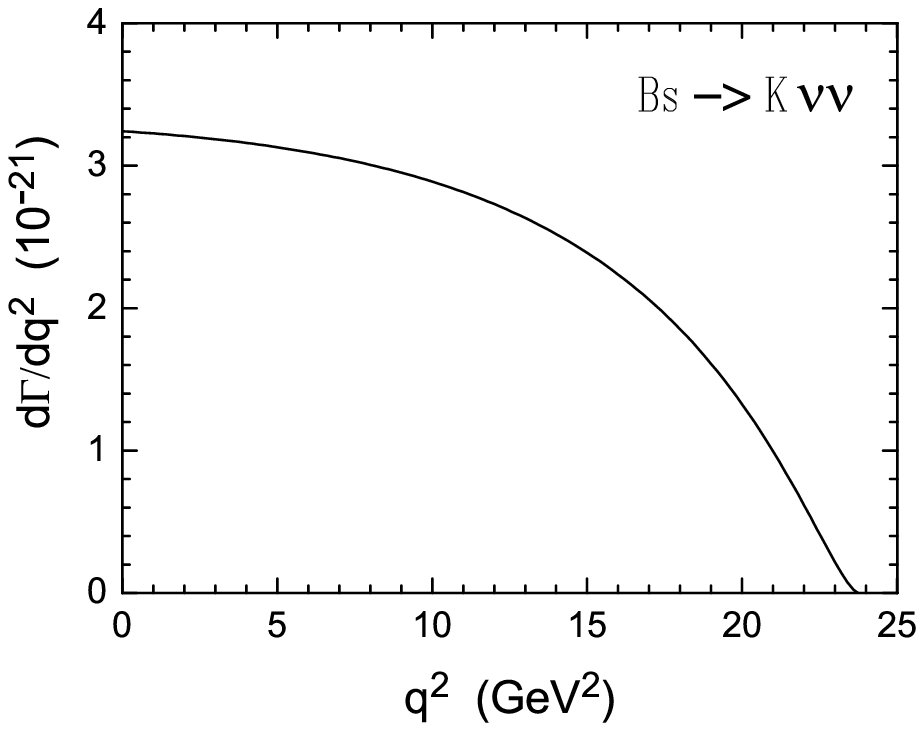}  }
\caption{The $q^2$-dependence of the differential decay rates $d\Gamma/dq^2$
for the decay processes with the $B_s\to K$ transitions.}
\label{fig:fig7}
\end{center}
\end{figure}

Firstly, in Figs.(\ref{fig:fig6}-\ref{fig:fig7}),  we show the differential
decay rates $d\Gamma/dq^2$ for the decay modes corresponding to the
$B \to (\pi, K)$ and $B_s \to K$ transitions.

By making the numerical integrations over the whole range of $q^2$, we find
the numerical results for the branching ratios.
For the $b \to u$ charged current processes, the NLO pQCD predictions for the decay rates
are the following
\beq
Br(\bar{B}^0 \to \pi^+ l^- \bar{\nu}_l) &=& \left (1.42
^{+0.40}_{-0.30}(\omega_b) \pm 0.15 (a_2^\pi) \pm 0.12 (m_0^\pi)  \right)
\times\frac{|V_{ub}|^2}{|0.0038|^2}\times 10^{-4},
\label{eq:brpilnu}\\
Br(\bar{B}^0 \to \pi^+ \tau^- \bar{\nu}_\tau) &=&\left (0.90
^{+0.25}_{-0.19}(\omega_b)^{+0.09}_{-0.08}(a_2^\pi)\pm 0.08 (m_0^\pi)  \right)
\times\frac{|V_{ub}|^2}{|0.0038|^2}\times 10^{-4},
\label{eq:br01}
\eeq
\beq
Br(B^- \to \pi^0 l^- \bar{\nu}_l) &=&\left (7.63
^{+2.16}_{-1.61}(\omega_b) ^{+0.82}_{-0.78}(a_2^\pi)^{+0.66}_{-0.63}(m_0^\pi)  \right)
\times\frac{|V_{ub}|^2}{|0.0038|^2}\times 10^{-5}, \label{eq:brpi0lnu}
\\
Br(B^- \to \pi^0 \tau^- \bar{\nu}_\tau) &=& \left (4.85
^{+1.34}_{-1.01}(\omega_b)^{+0.47}_{-0.41}(a_2^\pi)^{+0.45}_{-0.43}(m_0^\pi)  \right)
\times\frac{|V_{ub}|^2}{|0.0038|^2}\times 10^{-5},
\label{eq:br02}\\
Br(\bar{B}_s^0 \to K^+ l^- \bar{\nu}_l) &=& \left (1.27
^{+0.46}_{-0.26}(\omega_{bs})^{+0.14}_{-0.13}(a_i^K)^{+0.10}_{-0.09}(m_0^K)  \right)
\times\frac{|V_{ub}|^2}{|0.0038|^2}\times 10^{-4},\label{eq:br03}\\
Br(\bar{B}_s^0 \to K^+ \tau^- \bar{\nu}_\tau) &=& \left (7.78
^{+2.51}_{-1.81}(\omega_{bs}) ^{+0.69}_{-0.66}(a_i^K) ^{+0.62}_{-0.59} (m_0^K)  \right)
\times\frac{|V_{ub}|^2}{|0.0038|^2}\times 10^{-5},
\label{eq:br03b}
\eeq
where the first error comes from the uncertainties of $\omega_b=0.40\pm 0.04$ or
$\omega_{B_s}=0.50\pm 0.05$, the second error are induced by the variations of
$a_1^K=0.06\pm 0.03$, $a_2^{\pi,K}=0.25\pm 0.15$ and the third error comes from
the uncertainties of $m_0^\pi=1.4\pm 0.1GeV$,
$m_0^K=1.6\pm 0.1 GeV$.

For $\bar{B}^0 \to \pi^+ l^- \bar{\nu}_l$ and $B^- \to \pi^0 l^- \bar{\nu}_l$ decay mode, their branching ratios have been
well measured by BaBar, Belle and CLEO Collaborations
\cite{babar-83-032007,cleo-99-041802,belle-648-139}.
The new BaBar measurement\cite{babar-83-032007} and the new world average \cite{pdg2012}
are the following
\beq
Br\left (\bar{B}^0 \to \pi^+ l^- \bar{\nu}_l\right)&=&\left\{ \begin{array}{l}
\left ( 1.41\pm 0.05(syst.)
\pm 0.07(stat.) \right )\times 10^{-4},  \ \   {\rm BaBar} [1],  \\
\left ( 1.44\pm 0.05 \right )\times 10^{-4}, \ \   {\rm PDG2012} [37],  \\
\end{array} \right. \label{eq:brexp1}\\
Br(B^- \to \pi^0 l^- \bar{\nu}_l) &=&\left (7.78 \pm 0.28   \right) \times 10^{-5},
\ \   {\rm PDG2012} [37], \label{eq:brexp2}
\eeq
On the other hand, we know that one can extract out the magnitude of the CKM matrix
element $V_{ub}$ by comparing the theoretical prediction for $Br(B\to \pi l\nu)$
with the data.

Based on the measured partial branching fraction for $B \to \pi l \nu$
in the range of $0\leq q^2< 12 GeV^2$ and the most recent QCD light-cone sum-rule
calculation of the form factor $F_+(q^2)$\cite{jpcs-110052026},
BaBar Collaboration found the result \cite{babar-83-032007}
\beq
|V_{ub}| =\left ( 3.78 \pm 0.13(exp.) ^{+0.55}_{-0.40}(theor.)\right ) \times
10^{-3},
\label{eq:vub1}
\eeq
where the two errors refer to the experimental and theoretical uncertainties.

From the differential decay rate  as given in Eq.(\ref{eq:dfdq2}) and the pQCD calculation  of the form  factor
$F_+(q^2)$ at the NLO level, we make the numerical integration over
the whole range of $0 \leq q^2 \leq (M_B-m_\pi)^2$, compare the
obtained branching ratio for $\bar{B}^0\to \pi^+ l^- \bar{\nu}_l$ with
the new BaBar measurement as given in Eq.~(\ref{eq:brexp1}), and derive out our estimation for the magnitude of
$V_{ub}$:
\beq
|V_{ub}| =\left ( 3.80^{+0.50}_{-0.43}(\omega_b)\pm 0.20(a_2^\pi) ^{+0.16}_{-0.15}(m_0^\pi)\right ) \times10^{-3}
=\left (3.80^{+0.56}_{-0.50}(theor.) \right)\times 10^{-3},
\label{eq:vub2}
\eeq
It is easy to see that our estimation for the central value and the uncertainty of $|V_{ub}|$ agrees very well
with the BaBar result as given in Eq.~(\ref{eq:vub1}).

For other neutral current processes, after making the numerical integration over the whole range
of $0 \leq q^2 \leq (M_B-m_\pi)^2$, we find the NLO pQCD predictions for the branching ratios.
The pQCD predictions at the NLO level and currently available data are all listed in Table \ref{tab-brn1}.
The first error of the pQCD predictions comes from the uncertainties of $\omega_b$ or $\omega_{B_s}$, the
second one from $a_1^K$ and/or $a_2^{\pi,K}$, and the third one is induced by the uncertainties of the
chiral mass $m_0^{\pi,K}$.


\begin{table}[thb]
\begin{center}
\caption{The NLO pQCD predictions for the branching ratios of the considered
decays with $l=(e,\mu)$ and currently available experimental measurements
\cite{babar-83-032007,cleo-99-041802,belle-648-139,babar-1204-3933,belle-0410006} and the world averages\cite{pdg2012}.
The upper limits are all given at the $90\%$C.L.}
\label{tab-brn1}
\begin{tabular}{l| l |l}  \hline\hline
 Decay modes       & \hspace{0.5cm}NLO pQCD predictions \hspace{0.5cm}& \hspace{0.5cm} Data \\  \hline
$Br(\bar{B}^0\to\pi^0 l^+ l^-)$  & $\left (0.91^{+0.26}_{-0.19} \pm 0.10 \pm 0.08 \right )\times 10^{-8}$ & $< 1.2\times 10^{-7}$  \\
$Br(\bar{B}^0\to\pi^0 \tau^+ \tau^-)$  & $\left (0.28^{+0.07}_{-0.06}\pm 0.02\pm 0.03 \right )\times 10^{-8}$ &   \\
$Br(\bar{B}^0\to\pi^0 \nu  \bar{\nu})$ & $\left (7.30^{+2.07+0.79+0.63 }_{-1.54-0.74-0.61}\right )\times 10^{-8} $ & $<2.2\times 10^{-4}$  \\ \hline
$Br(B^-\to\pi^- l^+ l^-)$  & $\left (1.95^{+0.55+0.21+0.17}_{-0.41-0.20-0.16} \right )\times 10^{-8}$ & $< 4.9 \times 10^{-8}$  \\
$Br(B^-\to\pi^- \tau^+ \tau^-)$  & $\left (0.60^{+0.16+0.04}_{-0.12-0.03} \pm 0.06\right )\times 10^{-8}$ &   \\
$Br(B^-\to\pi^- \nu  \bar{\nu})$ & $\left (1.57^{+0.44+0.17+0.14}_{-0.33-0.16-0.13}\right )\times 10^{-7}$ & $< 1.0\times 10^{-4}$  \\ \hline
$Br(\bar{B}^0\to\bar{K}^0 l^+ l^-)$  & $\left (5.1^{+1.5+0.5+0.4}_{-1.1-0.5-0.4}\right )\times 10^{-7}$ & $(4.7^{+0.6}_{-0.2})\times 10^{-7}$  \\
$Br(\bar{B}^0\to\bar{K}^0 \tau^+ \tau^-)$  & $\left (1.20^{+0.32+0.07+0.11}_{-0.25-0.07-0.10}\right )\times 10^{-7}$ &   \\
$Br(\bar{B}^0\to\bar{K}^0 \nu  \bar{\nu})$ & $\left (4.1^{+1.2+0.4+0.3}_{-0.9-0.4-0.3}\right )\times 10^{-6}$ &
$ < 5.6\times 10^{-5} $  \\ \hline
$Br(B^-\to K^- l^+ l^-)$  & $\left (5.50^{+1.59+0.57+0.42}_{-1.18-0.55-0.41}\right )\times 10^{-7}$ &  $(5.1\pm 0.5)\times 10^{-7}$  \\
$Br(B^-\to K^- \tau^+ \tau^-)$  & $ \left (1.29^{+0.35}_{-0.26} \pm 0.08\pm 0.11 \right )\times 10^{-7}$ &   \\
$Br(B^-\to K^-  \nu  \bar{\nu})$ & $ \left (4.42^{+1.28+0.46+0.34}_{-0.95-0.44-0.33}\right )\times 10^{-6}$ &  $< 1.3\times 10^{-5}$  \\ \hline
$Br(\bar{B}_s^0\to K^0 l^+ l^-)$  & $\left (1.633^{+0.54+0.18+0.12}_{-0.38-0.17-0.12}\right )\times 10^{-8}$ &   \\
$Br(\bar{B}_s^0\to K^0 \tau^+ \tau^-)$  & $\left (0.43^{+0.13+0.03+0.04}_{-0.10-0.03-0.04}\right )\times 10^{-8}$ &   \\
$Br(\bar{B}_s^0\to K^0 \nu  \bar{\nu})$ & $\left (1.31^{+0.43+0.14+0.10}_{-0.31-0.13-0.10}\right )\times 10^{-7}$ &   \\ \hline
\hline
 \end{tabular}
 \end{center}
\end{table}

From the NLO pQCD predictions for the branching ratios of all considered
semi-leptonic decays of $B$ and $B_s$ meson, as listed in
Eqs.~(\ref{eq:br01}-\ref{eq:br03}) and Table \ref{tab-brn1}, we have the
following points:
\begin{enumerate}
\item[(i)]
The branching ratios of the charged current processes $B\to\pi l\nu$ and $B_s\to Kl\nu$ are all at
the order of $10^{-4}$.
For $\bar{B}^0 \to \pi^+l^-\bar{\nu}_l$ and $B^- \to \pi^0 l^-\bar{\nu}_l$  decay modes, the
pQCD predictions for its branching ratios as shown in Eqs.~(\ref{eq:brpilnu},\ref{eq:brpi0lnu})
agree very well with the data as given in Eqs.~(\ref{eq:brexp1},\ref{eq:brexp2}). For other
charged current decay modes, the pQCD predictions as given in
Eqs.~(\ref{eq:br01},\ref{eq:br02}-\ref{eq:br03b}) will be tested by the LHCb and the forthcoming
Super-B experiments.

\item[(ii)]
For the neutral current $\bar{B}^0\to\bar{K}^0 l^+ l^-$ and $B^- \to K^- l^+ l^-$ decays,
the NLO pQCD predictions for their branching ratios agree very well with currently available
experimental measurements. For other neutral current decays, the NLO pQCD predictions are all
consistent with currently available experimental upper limits and will be tested by LHCb
and the fothcoming Super-B experiments.

\item[(iii)]
Because of the strong suppression of the  CKM factor $|V_{td}/V_{ts}|^2=0.211^2$
~\cite{pdg2010}, the branching ratios for the decays with $b\to d$ transitions
are much smaller than those decays with the $b\to s$ transitions.
Furthermore, the branching ratios of $B_{(s)}\to P\nu\nu$ are almost an order larger
than their corresponding decay modes $B_{(s)}\to P l^+ l^- $
partially due to the generation factor $N_g=3$.
In order to reduce the theoretical uncertainty of the pQCD predictions, we defined several ratios
$R_\nu,R_C$ and $R_{N1,N2,N3}$ among the branching ratios of the considered decay modes.

The NLO pQCD prediction for the ratio $R_\nu $ is of the form
\beq
R_\nu &=& \frac{Br(\bar{B}^0 \to \pi^0 \nu \bar{\nu}) }{
Br(\bar{B}^0 \to \pi^0  l^+ l^-) }\approx
\frac{Br(\bar{B}^0 \to \bar{K}^0 \nu \bar{\nu}) }{
Br(\bar{B}^0 \to \bar{K}^0  l^+ l^-) }\non
& \approx &
 \frac{Br(B^- \to \pi^- \nu \bar{\nu}) }{
Br(B^- \to \pi^-  l^+ l^-) }\approx
\frac{Br(B^- \to K^- \nu \bar{\nu}) }{
Br(B^- \to K^-  l^+ l^-) }\non
&\approx&  \frac{Br(\bar{B}_s^0 \to K^0 \nu \bar{\nu}) }{
Br(\bar{B}^0_{s} \to K^0 l^+ l^-)} \approx 8,
\label{eq:rnu01}
\eeq
for $l=(e,\mu)$. These relations will be tested by experiments.

\item[(iv)]
Because of the large mass of $\tau$ lepton,  we found that the considered $B/B_s$ decays involving one or two $\tau$'s in the
final state have a smaller decay rates than those without $\tau$. The pQCD predictions for the ratios $R_C$ and $R_{N1,N2,N3}$
of the corresponding branching ratios of relevant decays are the following
\beq
R_C &=& \frac{Br(\bar{B}^0_{s})\to P^+ l^- \bar{\nu}_l) }{Br(\bar{B}^0_{s})\to P^+ \tau^- \bar{\nu}_\tau) }\approx 1.5,
\label{eq:rc01}\\
R_{N1} &=& \frac{Br(\bar{B}^0 \to \pi^0 l^+ l^-) }{
Br(\bar{B}^0 \to \pi^0 \tau^+ \tau^-) }\approx
\frac{Br(B^- \to \pi^- l^+ l^-) }{
Br(B^- \to \pi^- \tau^+ \tau^-) } \approx 3.3,
\label{eq:rn01}\\
R_{N2} &=& \frac{Br(\bar{B}^0 \to \bar{K}^0 l^+ l^-) }{
Br(\bar{B}^0 \to \bar{K}^0 \tau^+ \tau^-) }\approx
\frac{Br(B^- \to K^- l^+ l^-) }{
Br(B^- \to K^- \tau^+ \tau^-) } \approx 4.3,
\label{eq:rn02}\\
R_{N3} &=& \frac{Br(\bar{B}^0_{s} \to K^0 l^+ l^-) }{
Br(\bar{B}^0_{s} \to K^0 \tau^+ \tau^-) }\approx 3.8,
\label{eq:rn03}
\eeq
for $l=(e,\mu)$. These relations will be tested by LHCb and the forthcoming Super-B experiments.

\end{enumerate}

\section{Summary and conclusions} \label{sec:5}

In this paper we calculated the branching ratios of the semileptonic decays $B \to (\pi,K)(l^+l^-, l\nu, \nu \bar{\nu})$
and $B_s \to K (l^+l^-, l\nu, \nu \bar{\nu})$ in the pQCD factorization approach.
We firstly evaluate the $B \to (\pi, K)$ and $B_s \to K$ transition form factors
$F_{0,+,\rmt}(q^2)$ by employing the pQCD factorization approach with the inclusion of the
next-to-leading-order corrections, and then we calculate the branching ratios for all considered semileptonic decays.
Based on the numerical results and the phenomenological analysis, we found the following points:
\begin{enumerate}
\item[]{(i)}
For the $B \to (\pi, K)$ and $B_s \to K$ transition form factors $F_{0,+,\rmt}(q^2)$, the
NLO pQCD predictions for the values of $F_{0,+,\rmt}(0)$ and their $q^2$-dependence agree well with those
obtained from the LCSR or other methods. The NLO part of the form factors in the pQCD factorization
approach is only around $20\%$ of the total value.

\item[]{(ii)}
For the charged current $\bar{B}^0 \to \pi^+l^-\bar{\nu}_l$ and $B^- \to \pi^0 l^-\bar{\nu}_l$  decays and
the neutral current $\bar{B}^0\to\bar{K}^0 l^+ l^-$ and $B^- \to K^- l^+ l^-$ decays, the
NLO pQCD predictions for their branching ratios agree very well with the measured values.

\item[]{(iii)}
By comparing the pQCD predictions for $Br(\bar{B}^0 \to \pi^+l^-\bar{\nu}_l)$ with the measured 
decay rate we extract out the magnitude of the CKM element  $V_{ub}$: 
$|V_{ub}|= \left ( 3.80^{+0.56}_{-0.50}(theor.)\right ) \times10^{-3}$.

\item[]{(iv)}
We also defined several ratios of the branching ratios $R_\nu, R_C$ and $R_{N1,N2,N3}$, and presented the corresponding
pQCD predictions, which will be tested by LHCb and the forthcoming Super-B experiments.

\end{enumerate}

\begin{acknowledgments}
The authors would like to thank H.n. Li, Y.M. Wang, C.D. Lu and
Y.L. Shen for valuable discussions.
This work is supported in part by the National Natural Science Foundation of
China under Grant No. 10975074 and 10735080.
\end{acknowledgments}


\appendix

\section{Related functions defined in the text}

In this Appendix, we present the functions needed in the pQCD calculation.
The threshold resummation factors $S_t(x)$ is adopt from Ref.\cite{li-65-014007}:
\beq
S_t=\frac{2^{1+2c}\Gamma(3/2+c)}{\sqrt{\pi}\Gamma(1+c)}[x(1-x)]^c,
\eeq
and we here set the  parameter $c=0.4$.
The hard functions $h_1$ and $h_2$ come form the Fourier transform and can be written as
\cite{li-83054029}
\beq
h_1(x_1,x_2,b_1,b_2)&=&K_0\left (\sqrt{x_1x_2\eta}m_Bb_1 \right )
\Bigl [\theta(b_1-b_2)I_0\left (\sqrt{x_2\eta}m_Bb_2\right )
K_0\left (\sqrt{x_2\eta}m_Bb_1 \right)\non
&& +\theta(b_2-b_1)I_0\left (\sqrt{x_2\eta}m_Bb_1 \right )
K_0\left (\sqrt{x_2\eta}m_Bb_2\right )\Bigr]S_t(x_2),
\eeq
\beq
h_2(x_1,x_2,b_1,b_2)&=&\frac{b_1 K_1\left (\sqrt{x_1x_2\eta}m_Bb_1\right )}{
2\sqrt{x_1x_2\eta}m_B}
\Bigl [\theta(b_1-b_2)I_0\left (\sqrt{x_2\eta}m_Bb_2 \right)
K_0\left (\sqrt{x_2\eta}m_Bb_1 \right ) \non 
&&+\theta(b_2-b_1) I_0\left (\sqrt{x_2\eta}m_Bb_1 \right )
K_0\left (\sqrt{x_2\eta}m_Bb_2 \right ) \Bigr] S_t(x_2),
\eeq
where $J_0$ is the Bessel function and $K_0$, $K_1$, $I_0$ are modified Bessel functions.

The factor $exp[-S_{ab}(t)]$ contains the Sudakov logarithmic corrections and
the renormalization group evolution effects of both the wave functions and the
hard scattering amplitude with $S_{ab}(t)=S_B(t)+S_P(t)$, where
\beq
S_B(t)&=&s\left (x_1\frac{m_B}{\sqrt{2}},b_1 \right )+\frac{5}{3}\int_{1/b_1}^{t}
\frac{d\bar{\mu}}{\bar{\mu}}\gamma_q\left (\alpha_s(\bar{\mu}) \right ),\\
S_P(t)&=&s\left (x_2\frac{m_B}{\sqrt{2}},b_2 \right )
+s\left ((1-x_2)\frac{m_B}{\sqrt{2}},b_2 \right )
+2\int_{1/b_2}^{t}\frac{d\bar{\mu}}{\bar{\mu}}\gamma_q\left (\alpha_s(\bar{\mu}) \right ),
\eeq
with the quark anomalous dimension $\gamma_q=-\alpha_s/\pi$.
The functions $s(Q,b)$ are defined by \cite{li-65-014007}
\beq
s(Q,b)&=&\frac{A^{(1)}}{2\beta_1}\hat{q}\ln\left (\frac{\hat{q}}{\hat{b}}\right)
-\frac{A^{(1)}}{2\beta_1}(\hat{q}-\hat{b})
+\frac{A^{(2)}}{4\beta_1^2}\left (\frac{\hat{q}}{\hat{b}}-1 \right )\non
&&-\left [\frac{A^{(2)}}{4\beta_1^2}-\frac{A^{(1)}}{4\beta_1}
\ln\left (\frac{e^{2\gamma_E}-1}{2}\right )\right ]
\ln\left ({\frac{\hat{q}}{\hat{b}}}\right )\non
&& +\frac{A^{(1)}\beta_2}{4\beta_1^3}\hat{q}\left [\frac{\ln(2\hat{q})+1}{\hat{q}}
-\frac{\ln(2\hat{b})+1}{\hat{b}} \right ]
+\frac{A^{(1)}\beta_2}{8\beta_1^3} \left [ \ln^2(2\hat{q})-\ln^2(2\hat{b})\right ],
\eeq
where the variables are defined by $ \hat{q}=\ln[Q/(\sqrt{2}\Lambda)]$, 
$\hat{b}=\ln[1/(b\Lambda)]$, and the coefficients $A^{(i)}$ and $\beta_i$ are
\beq
\beta_1&=&\frac{33-2n_f}{12}, \quad \beta_2=\frac{153-19n_f}{24},\quad
A^{(1)}=\frac{4}{3}, \notag \\
 \quad A^{(2)}&=&\frac{67}{9}-\frac{\pi^2}{3}-\frac{10n_f}{27}
+\frac{8}{3}\beta_1\ln(e^{\gamma_E}/2)
\eeq
here, $n_f$ is the number of the quark flavors, the $\gamma_E$ is the Euler constant.
The hard scales $t_i$ in the equations of this work are chosen as the largest scale of the
virtuality of the internal particles in the hard $b$-quark decay diagram,
\beq
t_1=\max\{\sqrt{x_2\eta}m_B,1/b_1,1/b_2\},  \quad
t_2=\max\{\sqrt{x_1\eta}m_B,1/b_1,1/b_2\}
\eeq


\end{document}